%% file: NIKA_MACSJ1424.tex
\documentclass[traditabstract]{aa}  

\usepackage{fixltx2e}
\usepackage[english]{babel}
\usepackage{graphicx,amsmath}
\usepackage{epstopdf}
\usepackage{epsf,color}
\usepackage[mathscr]{eucal}
\usepackage{amsmath}
\usepackage{amssymb,amsfonts}
\usepackage{natbib}
\usepackage{graphicx}
\usepackage{txfonts}
\usepackage{dsfont}
\definecolor{Mygreen}{rgb}{0.00, 0.72, 0.0}
\definecolor{Mypink}{rgb}{1.0, 0.0, 0.5}
\usepackage[breaklinks, citecolor=blue, linkcolor=Mygreen, urlcolor=Mypink, colorlinks=true, debug, baseurl=' ']{hyperref}
\usepackage{float}
\usepackage{color}

\def\simlt{\lower.5ex\hbox{$\; \buildrel < \over \sim \;$}}
\def\simgt{\lower.5ex\hbox{$\; \buildrel > \over \sim \;$}}

\bibpunct{(}{)}{;}{a}{}{,}
\newfont{\gwpfont}{cmssq8 scaled 1000}
\newcommand{\rexcess}{{\gwpfont REXCESS}}

\begin{document}

\title{High angular resolution Sunyaev-Zel'dovich observations of \mbox{MACS~J1423.8+2404} with NIKA: Multiwavelength analysis}
\input{listeauthors}
\date{Received 22 October 2015 \ / Accepted 16 December 2015}

\abstract {The prototype of the NIKA2 camera, NIKA, is a dual-band instrument operating at the IRAM 30-meter telescope, which can observe the sky simultaneously at 150 and 260~GHz. One of the main goals of NIKA (and NIKA2) is to measure the pressure distribution in galaxy clusters at high angular resolution using the thermal Sunyaev-Zel'dovich (tSZ) effect. Such observations have already proved to be an excellent probe of cluster pressure distributions even at intermediate and high redshifts. However, an important fraction of clusters host sub-millimeter and/or radio point sources, which can significantly affect the reconstructed signal. Here we report  on $<20$ arcsec angular resolution observations at 150 and 260~GHz of the cluster \mbox{MACS~J1423.8+2404}, which hosts both radio and sub-millimeter point sources. We examine the morphological distribution of the tSZ signal and compare it to other datasets. The NIKA data are combined with Herschel satellite data to study the spectral energy distribution (SED) of the sub-millimeter point source contaminants. We then perform a joint reconstruction of the intracluster medium (ICM) electronic pressure and density by combining NIKA, Planck, XMM-Newton, and Chandra data, focusing on the impact of the radio and sub-millimeter sources on the reconstructed pressure profile. We find that  large-scale pressure distribution is unaffected by the point sources because of the resolved nature of the NIKA observations. The reconstructed pressure in the inner region is slightly higher when the contribution of point sources are removed. We show that it is not possible to set strong constraints on the central pressure distribution without  accurately removing these contaminants. The comparison with X-ray only data shows good agreement for the pressure, temperature, and entropy profiles, which all indicate that \mbox{MACS~J1423.8+2404} is a dynamically relaxed cool core system. The present observations illustrate  the possibility of measuring these quantities with a relatively small integration time, even at high redshift and without X-ray spectroscopy. This work is part of a pilot study aiming at optimizing tSZ observations with the future NIKA2 camera.}

\titlerunning{Resolved tSZ observations of MACS~J1423.8+2404}
\authorrunning{R. Adam, B. Comis, I. Bartalucci et al.}
\maketitle

\section{Introduction}\label{sec:Introduction}
In the standard scenario of structure formation, clusters of galaxies form through the hierarchical merging of smaller groups and  accretion of surrounding material. These clusters are sensitive to both the matter content of the Universe and its dynamics because they form throughout its expansion history. Their formation process is  well understood and clusters have been used to place constraints on cosmological parameters \citep[e.g.,][]{planck2013cluster_count}. However, the complex baryonic physics occurring during cluster formation, such as feedback from active galactic nuclei (AGN), or nonthermal processes occurring during mergers or in the presence of intracluster medium (ICM) turbulence or coherent motion, is still unclear \citep[see for example][]{borgani2011}. This leads to scatter and biases in the observable--mass scaling relations needed to compare to theory, limiting the use of clusters as cosmological probes.

The thermal Sunyaev-Zel'dovich effect \citep[tSZ;][]{sunyaev1972,sunyaev1980} is produced by the inverse Compton interaction of cosmic microwave background (CMB) photons with the energetic electrons in the ICM. This tSZ effect leads to a spectral distortion of the CMB observable at millimeter and sub-millimeter wavelengths that is directly proportional to the line-of-sight integral of the electronic pressure distribution in the ICM. The integrated tSZ flux is related to the overall thermal energy of the cluster and is therefore expected to provide a low scatter mass proxy with a small dependence on the dynamical state of the cluster or the exact gas physics \citep[e.g.,][]{dasilva2004,motl2005,nagai2006}. Furthermore, resolved tSZ observations are very sensitive to the overpressure caused by mergers and have proved to be very efficient for probing cluster astrophysics \citep[see, for example, results by][]{pointecouteau1999,komatsu2001,korngut2011,adam2013,young2014,adam2014,mroczkowski2015}. This is particularly true at high redshifts since, unlike other probes, the tSZ signal does not suffer from cosmological dimming and is only limited by the sensitivity and angular resolution of the observations. Detailed reviews of the tSZ effect can be found in \cite{birkinshaw1999,carlstrom2002}, and \cite{kitayama2014}.

During the past few years, tremendous achievements have been made in the tSZ community with the production of catalogues of more than 1\,000 objects by Planck \citep{planck2013catalogue}, the Atacama Cosmology Telescope \citep[ACT;][]{hasselfield2013}, and the South Pole Telescope \citep[SPT;][]{reichardt2013,bleem2014}, at $>1$ arcmin angular resolution. X-ray observations have shown that, when  scaled to a characteristic radius, the cluster pressure profile shows a low dispersion \citep{arnaud2010}. This profile has now been well measured at intermediate scales in the nearby Universe with tSZ observations \citep{plagge2010,planck2013pressure_profile,sayers2013b}, but has not been deeply explored at high redshift and in the cluster cores because of the lack of high angular resolution observations. The investigation of this profile is nonetheless necessary to make better use of the available tSZ cluster samples when relating the tSZ signal to cluster mass.

In the context of high angular resolution tSZ observations, one of the main challenges to face is the removal of the contamination from point sources, as needed for example in the case of Bolocam observations \citep{sayers2013a}. Indeed, galaxy clusters contain galaxies that can host radio sources or a significant amount of dust. In addition, clusters at intermediate redshift provide optimal lenses, which can magnify sub-millimeter background galaxies \citep[see, e.g.,][]{adam2014}. Foreground galaxies can also be located in projection near the cluster under study. In tSZ observations, these objects appear as point-like contaminating sources.  The New IRAM Kids Array (NIKA) is the prototype of NIKA2, the next generation continuum instrument for the Institut de Radio Astronomie Millim\'etrique (IRAM) 30-meter telescope near Granada, Spain (see \citealt{monfardini2010,bourion2011,bourrion2012,monfardini2011,calvo2012,catalano2014}, for more details on the NIKA camera). The NIKA camera observes the sky at 150 and 260~GHz with an angular resolution of 18 and 12 arcsec full width at half maximum (FWHM), and has already been used to image the tSZ effect toward the galaxy clusters \mbox{RX~J1347.5-1145} and \mbox{CL~J1226.9+3332} \citep[see][]{adam2013,adam2014}. In the present paper we discuss observations of  the intermediate redshift cluster \mbox{MACS~J1423.8+2404} at $z~=~0.545$, which contains both radio and sub-millimeter galaxies and which we have used as a test case to investigate the impact of such contaminating objects on the reconstruction of the pressure of clusters observed with NIKA2. 

\mbox{MACS~J1423.8+2404} is a massive cluster from the MACS catalog \citep[Massive Cluster Survey;][]{ebeling2001}, for which a wealth of multiwavelength data have been obtained. \cite{schmidt2007}, in a study of relaxed clusters with Chandra, reported a virial mass $M_{\rm vir} = 4.52^{+0.79}_{-0.64} \times 10^{14}$ M$_{\odot}$, indicating that it indeed  is a massive object. The cluster was observed by the BIMA interferometer \citep{laroque2003} and by the Sunyaev-Zel'dovich Array (SZA), as part of a sample used to constrain the cluster pressure profile \citep{bonamente2012}. The large-scale structure of \mbox{MACS~J1423.8+2404} was investigated, using its red sequence galaxy density distribution \citep{kartaltepe2008}, showing a very relaxed morphology. \cite{guennou2014} studied the structure of the cluster using Chandra X-ray data as part of the dark energy American French team survey. They observed a strong X-ray emission, slightly elongated, and only low significance substructures were found. The morphology of the cluster was also studied from a detailed gravitational lensing/optical analysis by \cite{limousin2010}, who noticed that \mbox{MACS~J1423.8+2404} is nearly fully virialized, elongated, and shows very little substructure (see also the strong lensing results by \cite{zitrin2011}, as part of a sample of 12 MACS clusters). The temperature profile obtained by \cite{morandi2010} using Chandra data shows a typical cool core form with a low central temperature ($\sim 3$ keV) and a peak at $\sim 7$ keV at about 300 kpc away form the center. The brightest cluster galaxy (BCG) hosts a central AGN that is visible as a point source in radio observations \citep{condon1998,laroque2003,coble2007,bonamente2012}. This AGN is responsible for the presence of two cavities detected in the Chandra X-ray image \citep{hlavacek_larrondo2012}. Another radio source is located at about 1.5 arcmin southwest with respect to the X-ray peak. Cluster members, foreground, and background (including lensed sources) sub-millimeter galaxies are detected by Herschel, as observed during the Herschel Lensing Survey \citep[HLS;][]{egami2010,rawle2012}.

This paper is organized as follows. The observations performed at the IRAM 30-meter telescope are briefly presented in Sect.~\ref{sec:Observation_at_the_IRAM_30m_telescope_with_NIKA}. In Sect.~\ref{Radio_and_infrared_point_sources}, we discuss the processing of radio and sub-millimeter point sources and their impact on the radial tSZ flux density profile. The data reduction of the XMM-Newton and Chandra X-ray observations is presented in Sect.~\ref{sec:XMM_Newton_and_Chandra_X_ray_data_reduction}. We compare the NIKA observations to other multiwavelength datasets in Sect.~\ref{sec:Multi_wavelength_comparison}. In Sect. \ref{sec:Radial_pressure_reconstruction}, we reconstruct the radial pressure distribution of \mbox{MACS~J1423.8+2404} and explore the impact of the presence of the point sources. The pressure profile is combined with the electronic density of X-ray data to derive the thermodynamic distribution of the ICM. Conclusions and perspectives for NIKA2 are provided in Sect.~\ref{sec:conclusions}. Throughout this paper we assume a flat $\Lambda$CDM cosmology, according to the latest Planck results \citep{planck2014param}, with $H_0 = 67.8$ km s$^{-1}$ Mpc$^{-1}$, $\Omega_M = 0.308$, and $\Omega_{\Lambda} = 0.692$.

\section{Observations at the IRAM 30-meter telescope with NIKA}\label{sec:Observation_at_the_IRAM_30m_telescope_with_NIKA}
\subsection{The thermal Sunyaev-Zel'dovich effect}
The tSZ effect \citep{sunyaev1972,sunyaev1980} results in a distortion of the CMB blackbody spectrum relative to the primary CMB intensity, $I_0$, \citep[e.g.,][]{birkinshaw1999}
\begin{equation}
        \frac{\Delta I_{\rm tSZ}}{I_0} = y \ f(\nu, T_e).
\label{eq:deltaI}
\end{equation}
The function $f(\nu, T_e)$ gives the characteristic frequency dependence of the spectrum. The small dependance on the electronic temperature, $T_e$, arises from relativistic corrections for which we use the results of \cite{itoh1998}. The Compton parameter, $y$, gives the amplitude of the distortion and is related to the line-of-sight integral of the electronic pressure, $P_e$, as 
\begin{equation}
        y = \frac{\sigma_{\mathrm{T}}}{m_{e} c^2} \int P_{e} dl.
        \label{eq:y_compton}
\end{equation}
The parameter $\sigma_{\mathrm{T}}$ is the Thomson cross section, $m_{e}$ is the electron rest mass, and $c$ the speed of light. The total integrated tSZ flux, $Y_{\rm tot}$, is then given by the aperture photometry performed on the Compton parameter map.

In the NIKA bands, the tSZ signal is expected to be faint ($y \sim 10^{-4}$ for typical massive clusters) and diffuse. It is negative at 150~GHz and positive at 260~GHz.

\subsection{Observing conditions, scanning strategy, calibration, and data reduction}
\mbox{MACS~J1423.8+2404} was observed during the first NIKA open pool in February 2014. We collected 1.47 hours of on-target data. The atmospheric conditions were stable and the mean opacity was measured to be 0.14 and 0.15, at the source location, at 150 and 260~GHz, respectively, as detailed in \cite{catalano2014}. The mean elevation of the source was 30.8 degrees.

The scanning strategy adopted was the same as that used for \mbox{CL~J1226.9+3332}, detailed in \cite{adam2014}. Briefly, each scan consisted 19 subscans of 6 arcmin length separated by 10 arcsec steps made alternately at constant azimuth and constant elevation (relative to the map center). The pointing center was chosen to be at (R.A., Dec. 2000) = (14:23:47.8, +24:04:40.0) based on the Archive of Chandra Cluster Entropy Profile Tables catalog \citep[ACCEPT;][]{cavagnolo2009}.

The detailed calibration procedure can be found in \cite{adam2014} and we only summarize the main results here. Uranus was used as the primary calibrator. The Gaussian beam FWHM was measured to be 18.2 and 12.0 arcsec at 150 and 260~GHz, respectively. The nearby quasar 1354+195 was used to correct the pointing, for which the error is estimated to be less than 3 arcsec. The overall calibration uncertainty was estimated to be 7\% at 150~GHz and 12\% at 260~GHz. The NIKA bandpasses were used to convert the flux surface brightness to Compton parameter. We obtained $-10.9 \pm 0.8$ and $3.5 \pm 0.5$ Jy/beam per unit of Compton parameter at 150 and 260~GHz, respectively. The effective number of detectors was 117 at 150~GHz and 136 at 260~GHz, corresponding to an instantaneous field of view of 1.9 and 1.8 arcmin, respectively.

The removal of the atmospheric and electronic correlated noise, consisting of the subtraction of the correlated signal in the timelines across the detector arrays, was performed as in \cite{adam2014}. The resulting signal filtering was estimated using simulations. The observed transfer function was flat and close to unity at scales that are smaller than the field of view. The transfer function vanishes smoothly at larger angular scales \citep[see][]{adam2014}.

\subsection{Raw NIKA observations}\label{sec:Raw_NIKA_observations}
\begin{figure*}[h]
\centering
\includegraphics[height=6.6cm]{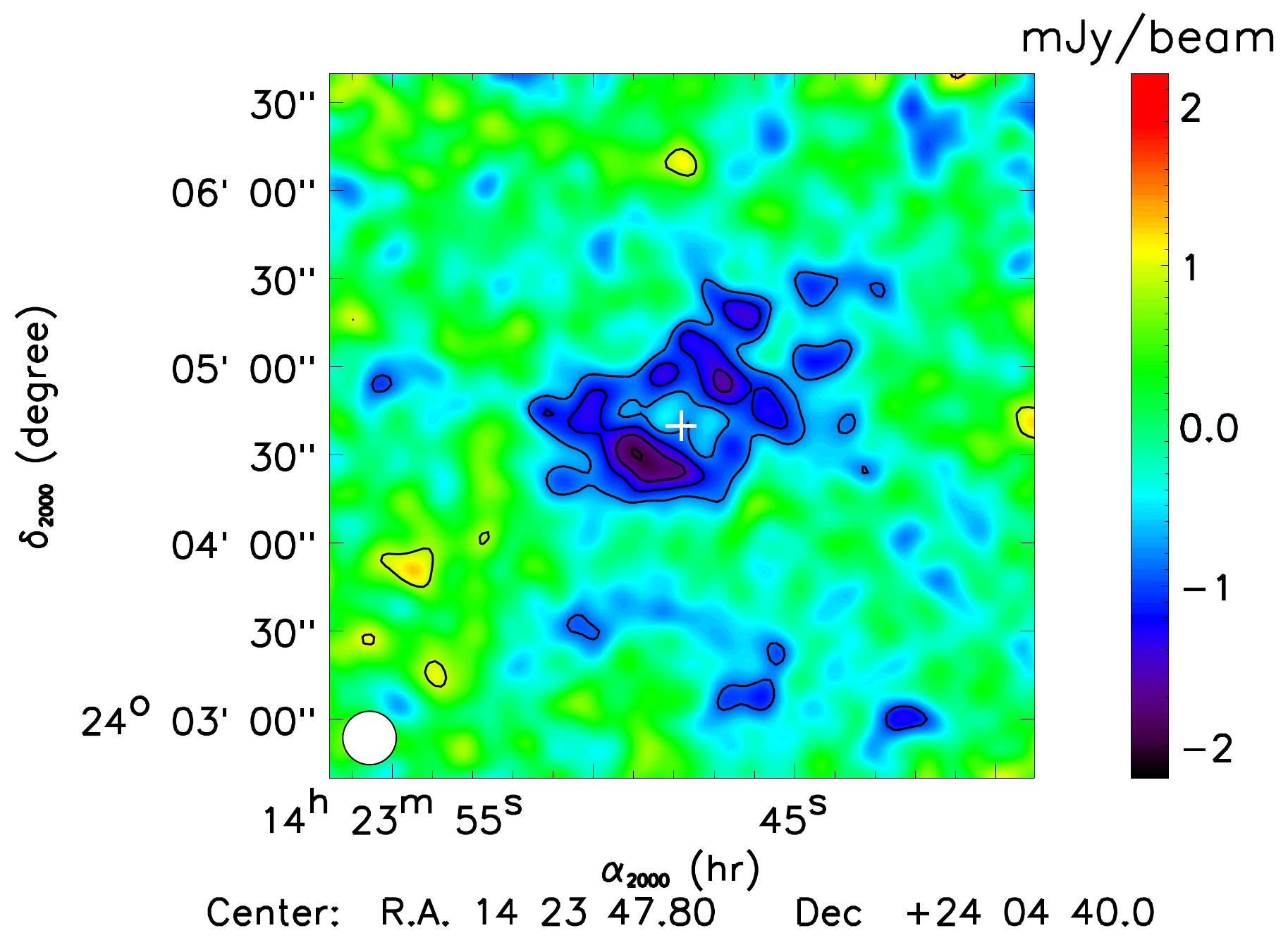}
\includegraphics[height=6.6cm]{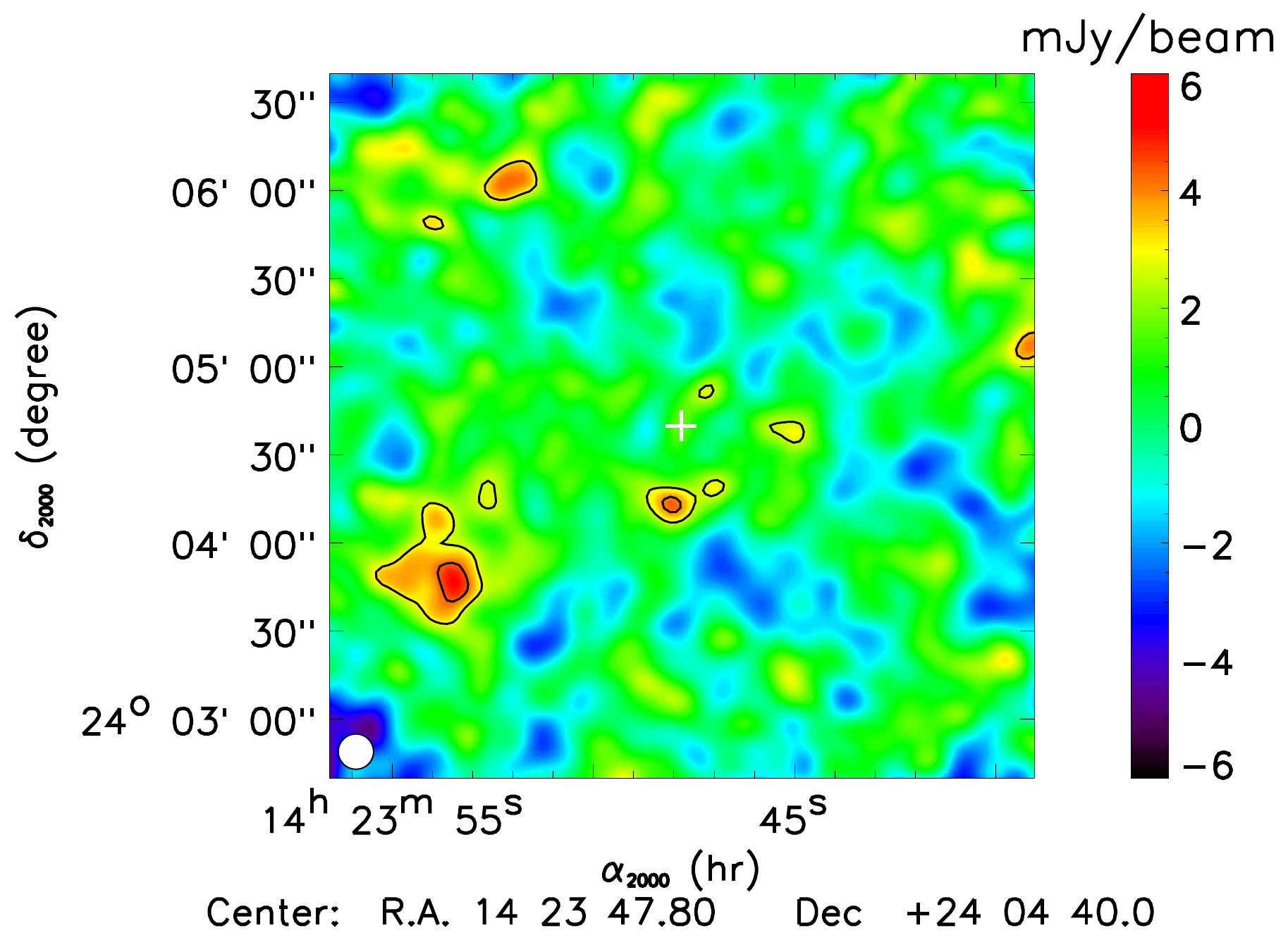}
\caption{{\footnotesize NIKA maps at 150~GHz (left) and 260~GHz (right) in units of surface brightness. The significance is given by the black contours starting at $\pm 2 \ \sigma$ with 1 $\sigma$ spacing. The maps are smoothed with an extra 10 arcsec Gaussian filter for display purposes and the effective beam FWHM is represented as a white circle in the bottom left corner of each panel. The white crosses indicate the X-ray center.}}
\label{fig:flux_map}
\end{figure*}
The NIKA maps that we obtained are presented in Fig.~\ref{fig:flux_map}. The 150~GHz map reveals a negative decrement, as expected from the tSZ effect at this frequency, with a maximum significance of $4.5 \ \sigma$. The morphology of the signal has a ring-like shape, as a consequence of the contamination by the radio point source at the cluster center, which fills in the tSZ signal. The 260~GHz map does not show any significant tSZ signal. At this frequency, extrapolating from the 150~GHz map, we expect a peak of $\sim 1$ mJy/beam, which is below the level of the noise standard deviation. We observe a $3 \ \sigma$ positive peak around (R.A., Dec.) = (14:23:53, +24:03:45), which corresponds to a $2 \ \sigma$ positive peak in the 150~GHz map owing to the presence of a sub-millimeter source. Another peak is also seen around (R.A., Dec.) = (+14:23:48, +24:04:15). The radio and sub-millimeter point source contamination is discussed in detail in Sect.~\ref{Radio_and_infrared_point_sources}.

The significance of the NIKA maps was calculated using Monte Carlo realizations. First, a noise map was obtained from the half difference of two equivalent subsets. After having normalized the noise map by the integration time per pixel, the noise spectral distribution was computed with the {\tt POKER} software \citep{ponthieu2011}, which properly accounts for incomplete sky coverage due to the scanning around the cluster. The noise power spectrum was modeled as a function of angular scale $k$, as $P_{\rm noise}(k) = A_{\rm white} + A_{\rm cor}(k_0) \left(\frac{k}{k_0}\right)^{\beta}$. The parameter $A_{\rm white}$ represents the intrinsic detector noise and $A_{\rm cor}$ and $\beta$ give the amplitude and the slope of the residual detector--detector atmospheric correlations in the map. This model was used to generate Monte-Carlo noise map realizations, $n_i$, accounting for the integration time per pixel. The noise realizations were smoothed by the same 10~arcsec Gaussian filter used to display the cluster, and the standard deviation across the Monte-Carlo realizations allowed us to compute the root mean square map and, therefore, the signal-to-noise contours shown in Fig.~\ref{fig:flux_map}. The noise standard deviation on the flux of point sources at the center of the map is 0.8 and 2.9 mJy at 150 and 260 GHz, accounting for reduction filtering effects. The mean noise standard deviation of the displayed maps within a radius of 60 arcsec, after applying the Gaussian smoothing, is 0.39 and 1.3 mJy/beam at 150 and 260~GHz, respectively. The full noise covariance matrix was also computed as the mean of the noise covariance over all the realizations, $C = \frac{1}{N_{\rm MC}}\sum_{i=1}^{N_{\rm MC}}n_i n_i^{T}$, and it is used in the analysis as described in Sect.~\ref{sec:Radial_pressure_reconstruction}.

\subsection{Astrophysical contaminants}\label{sec:astrophysical_contaminant}
In addition to the instrument noise and atmospheric noise residuals, diffuse astrophysical emission may give rise to an extra source of noise that cannot be reduced by increasing the observing time. In the following, we consider the contamination from CMB and cosmic infrared background (CIB) anisotropies, and Galactic emissions (dust, synchrotron, and free-free). 

The Galactic latitude of \mbox{MACS~J1423.8+2404} is high ($68.99^{\rm o}$), implying  low contamination of the cluster field by Galactic emissions \citep{planck2015X}. To estimate this contamination, we use the HEALPix software \citep{gorski2005} to extract 60 arcmin $\times$ 60 arcmin patches, centered on the cluster coordinates, of the nine Planck channel maps at 30, 44, 70, 100, 143, 217, 353, 547, and 857 GHz \citep{planck2015I}, to which we subtract the CMB map \citep{planck2015IX}. We find that CMB emission dominates over the low frequency Galactic components, i.e., synchrotron and free-free, at all frequencies. The CMB also dominates over the thermal dust emission, at higher frequencies, up to 353 GHz, where the amplitude of the fluctuations of the two are similar. In the considered field, the CMB therefore dominates over the diffuse Galactic emissions in the NIKA bands at the Planck angular resolution. This is be even more true at the smaller scales probed by NIKA as Galactic diffuse emissions have power spectra that are strongly decreasing with the spatial scale (power law with an index of about -2.5).

To estimate the level of contamination arising from CMB fluctuations, we use the CAMB software \citep{lewis2000} to compute the CMB power spectrum with the latest Planck cosmological parameters \citep{planck2014param} as input. The spectrum is computed up to multipole $\ell = 44000$ (corresponding to the NIKA angular resolution) and serves to generate CMB realization via the {\tt POKER} software \citep{ponthieu2011}. The simulated CMB maps are produced on a 5 arcmin $\times$ 5 arcmin field and are convolved with the NIKA transfer function including the beam smoothing. We find that the CMB fluctuations are about $15 \times 10^{-3}$ and $7 \times 10^{-3}$ mJy/beam, which is negligible with respect to the instrument noise and atmospheric noise residual at 150 and 260 GHz, respectively. Therefore, we neglect the CMB fluctuations and all the Galactic diffuse emission that are themselves fainter than the CMB fluctuations.

Finally, we estimate the contribution from extragalactic sources accounting for the clustering of dusty star-forming galaxies (CIB clustered anisotropies) and the shot noise from both dusty star-forming galaxies and radio sources. The clustering from radio sources is negligible \citep{hall2010}. The clustering term is computed using the CIB power spectrum measured at 143 and 217 GHz by \cite{planck2014XXX}, modeled by a 1-halo and a 2-halo term, and extrapolated to the NIKA frequencies. The shot noise, arising from unresolved sources below the NIKA detection threshold, is computed with the model from \cite{bethermin2012} in the case of dusty star-forming galaxies and \cite{tucci2011} for radio sources. At the considered scales, the shot noise dominates by a factor of about 5 with respect to the clustering component. Taking  the NIKA transfer function and beams into
account, we obtain a standard deviation of the fluctuations of 0.13 and 0.38 mJy/beam at 150 and 260 GHz in the NIKA bands. This extra source of noise is therefore subdominant with respect to the instrument noise and atmospheric noise residual at both frequencies, contributing to about 1\% of the noise when summed quadratically;  we do not account for the contribution of this extra noise source in the present paper.

\section{Radio and sub-millimeter point sources}\label{Radio_and_infrared_point_sources}
\subsection{Radio point sources}
 Very Large Array (VLA) data  at 4.8~GHz were used by \cite{laroque2003} to detect radio sources toward \mbox{MACS~J1423.8+2404}. Two sources were detected within the NIKA field and we used the coordinates obtained by \cite{laroque2003} as a reference (see Sect.~\ref{sec:Multi_wavelength_comparison} and Fig.~\ref{fig:MACSJ1424_mutiw}). The first source, hereafter RS1, is located within the central BCG, near the X-ray center. The second, hereafter RS2, is located at about 1.5 arcmin toward the southwest. The flux of RS1 was measured at various radio wavelengths between 1.4~GHz and 30~GHz, while the flux of RS2 was only measured at 1.4 and 4.8~GHz. We list in Table~\ref{tab:Radio_ps} the fluxes measured for both sources and the corresponding references. 
\begin{table*}[h]
\caption{{\footnotesize Location and flux of the radio sources observed in the $4 \times 4$ arcmin$^2$ field around \mbox{MACS~J1423.8+2404}.}}
\begin{center}
\begin{tabular}{ccccccc}
\hline
\hline
Source & Identifier & Position & 1.4 GHz & 4.8 GHz & 28.5 GHz & 30 GHz \\
 &  & [mJy] & [mJy] & [mJy] & [mJy] \\
\hline
RS1 & NVSS J142347+240439 & 14:23:47.78 +24:04:42.8$^{(a)}$ & $8.0 \pm 1.1 ^{(b)}$ & $4.40 \pm 0.03 ^{(a)}$ & $1.49 \pm 0.12 ^{(c)}$ & $2.0 \pm 0.2 ^{(d)}$ \\
RS2 & NVSS J142345+240340 & 14:23:45.07 +24:03:42.7$^{(a)}$ & $7.2 \pm 0.5 ^{(b)}$ & $2.72 \pm 0.03 ^{(a)}$ &  -- & --  \\  
\hline
\end{tabular}
\end{center}
{\small {\bf Notes.} $^{(a)}$ VLA; \cite{laroque2003}. $^{(b)}$ NVSS; \cite{condon1998}. $^{(c)}$ OVRO/BIMA; \cite{coble2007}. $^{(d)}$ SZA; \cite{bonamente2012}.}
\label{tab:Radio_ps}
\end{table*}

To estimate the expected flux of each source in the NIKA bands, we modeled their spectral energy distribution (SED) by $F_{\nu} = F_{1 \ {\rm GHz}} \left(\frac{\nu}{1 \ {\rm GHz}}\right)^{\alpha_{\rm radio}}$. The fluxes reported in Table~\ref{tab:Radio_ps} were used to fit the amplitude of the SED, $F_{1 \ {\rm GHz}}$, and its slope, $\alpha_{\rm radio}$. The best-fit parameters are reported in Table~\ref{tab:Radio_ps2} for both sources. We then simulated mock SEDs by sampling the parameters within their error bars and accounting for the covariance between them. Each mock SED was then integrated within the NIKA bandpasses to predict the expected flux at 150 and 260~GHz. The histogram of all the realizations was fitted with a Gaussian function to give the expected fluxes and uncertainties, which are listed for both sources and both NIKA frequencies in Table~\ref{tab:Radio_ps2}. The values of the spectral index $\alpha_{\rm radio}$ we obtain are typical for radio sources \citep[see for example][]{witzel1979}.
\begin{table*}[h]
\caption{\footnotesize Best-fit parameters and extrapolation of the fluxes in the NIKA bands of the radio sources in the $4 \times 4$ arcmin$^2$ field around \mbox{MACS~J1423.8+2404}. The degeneracy between the slope $\alpha_{\rm radio}$ and the amplitude $F_{1 \ {\rm GHz}}$ has been accounted for to extrapolate the flux in the NIKA bands. See text for details.}
\begin{center}
\begin{tabular}{ccccccc}
\hline
\hline
Source & R.A. offset & Dec. offset (arcsec) & $F_{1 \ {\rm GHz}}$ & $\alpha_{\rm radio}$ & 150 GHz & 260 GHz \\
 & [arcsec] & [arcsec] & [mJy] & & [mJy] & [mJy] \\
\hline
RS1 &      0.3 &      2.6 & $   10.39 \pm     0.30$ & $  -0.548 \pm    0.001$ & $    0.68 \pm     0.08$ & $    0.54 \pm     0.07$ \\
RS2 &     41.0 &    -57.3 & $    9.39 \pm     0.69$ & $  -0.790 \pm    0.003$ & $    0.18 \pm     0.04$ & $    0.13 \pm     0.03$ \\
\hline
\end{tabular}
\end{center}
\label{tab:Radio_ps2}
\end{table*}

\begin{table*}[h]
\caption{\footnotesize Positions and fluxes of the 17 sub-millimeter sources identified in the $4 \times 4$ arcmin$^2$ field around \mbox{MACS~J1423.8+2404}, measured by fitting Gaussian models to the maps at each wavelength as described in Sect.~\ref{sec:smmps}. The final two columns correspond to the NIKA bands.}
\begin{center}
\begin{tabular}{ccccccccc}
\hline
\hline
Source & 250 $\mu$m source position & 100 $\mu$m & 160 $\mu$m & 250 $\mu$m & 350 $\mu$m & 500 $\mu$m & 1.15 mm & 2.05 mm \\
 &  & [mJy] & [mJy] & [mJy] & [mJy] & [mJy] & [mJy] & [mJy] \\
\hline
SMG01 & 14:23:52.31 +24:05:04.9 & $    20.8 \pm      1.1$ & $    35.1 \pm      3.2$ & $    52.3 \pm      7.7$ & $    30.4 \pm      8.1$ & $    11.6 \pm      7.4$ & $     0.6 \pm      3.2$ & $     1.4 \pm      0.9$ \\
SMG02 & 14:23:48.16 +24:04:20.0 & $    12.4 \pm      1.4$ & $    21.0 \pm      3.1$ & $    35.8 \pm      9.8$ & $    24.1 \pm      7.4$ & $    16.3 \pm      7.2$ & $     4.8 \pm      2.9$ & $^{**}$ \\
SMG03 & 14:23:53.50 +24:06:05.1 & $    10.7 \pm      1.2$ & $    17.0 \pm      2.9$ & $    34.7 \pm     11.3$ & $    23.7 \pm      7.9$ & $     8.2 \pm      7.1$ & $     3.4 \pm      3.8$ & $     0.5 \pm      1.1$ \\
SMG04 & 14:23:42.42 +24:04:38.8 & $    11.5 \pm      1.4$ & $    17.7 \pm      3.0$ & $    29.1 \pm     10.8$ & $    21.3 \pm      8.0$ & $     4.4 \pm      7.5$ & $     2.6 \pm      3.2$ & $     0.4 \pm      0.9$ \\
SMG05 & 14:23:47.58 +24:04:48.7 & $     6.1 \pm      1.3$ & $     9.4 \pm      2.8$ & $    25.6 \pm      8.4$ & $    18.3 \pm      7.6$ & $    14.3 \pm      7.5$ & $     3.1 \pm      2.9$ & $^{**}$ \\
SMG06 & 14:23:53.32 +24:03:48.5 & $     2.9 \pm      1.4$ & $     8.5 \pm      3.1$ & $    20.6 \pm      9.6$ & $    14.9 \pm      8.0$ & $     1.8 \pm      7.7$ & $     8.2 \pm      3.4$ & $     1.0 \pm      0.9$ \\
SMG07 & 14:23:45.04 +24:05:48.9 & $     3.4 \pm      1.3$ & $    10.2 \pm      3.0$ & $    20.1 \pm      9.2$ & $    14.0 \pm      8.0$ & $     7.3 \pm      8.0$ & $     1.3 \pm      3.4$ & $     1.0 \pm      0.9$ \\
SMG08 & 14:23:49.16 +24:02:46.1 & $    -1.0 \pm      1.5$ & $     5.4 \pm      3.2$ & $    19.9 \pm      8.8$ & $    24.0 \pm      7.8$ & $    11.2 \pm      7.8$ & $     3.5 \pm      3.8$ & $     0.9 \pm      1.0$ \\
SMG09 & 14:23:43.27 +24:02:50.2 & $     2.4 \pm      1.3$ & $     5.8 \pm      3.3$ & $    12.6 \pm      8.6$ & $    13.8 \pm      7.7$ & $    19.4 \pm      7.8$ & $     2.5 \pm      4.1$ & $    -0.1 \pm      1.1$ \\
SMG10 & 14:23:44.55 +24:03:18.4 & $    -1.5 \pm      1.2$ & $     4.4 \pm      3.3$ & $    11.1 \pm      8.2$ & $    19.0 \pm      7.7$ & $    19.7 \pm      7.8$ & $     4.3 \pm      3.4$ & $    -0.2 \pm      0.9$ \\
SMG11 & 14:23:54.14 +24:05:32.2 & $     7.3 \pm      1.3$ & $    11.6 \pm      2.9$ & $    11.0 \pm      9.1$ & $     3.5 \pm      7.5$ & $     5.9 \pm      7.8$ & $    -1.6 \pm      3.7$ & $     0.3 \pm      1.0$ \\
SMG12 & 14:23:43.41 +24:03:50.6 & $     5.4 \pm      1.3$ & $     9.1 \pm      3.2$ & $     7.3 \pm     10.0$ & $    -1.9 \pm      7.6$ & $    -6.9 \pm      7.4$ & $     2.3 \pm      3.4$ & $     0.4 \pm      0.9$ \\
SMG13 & 14:23:50.13 +24:06:17.4 & $    11.7 \pm      1.3$ & $     8.9 \pm      3.1$ & $     6.4 \pm      7.5$ & $    -2.4 \pm      8.0$ & $   -10.5 \pm      7.7$ & $    -4.2 \pm      3.7$ & $     0.9 \pm      1.0$ \\
SMG14 & 14:23:47.36 +24:05:49.6 & $     3.9 \pm      1.3$ & $     5.5 \pm      2.9$ & $     7.9 \pm      8.6$ & $    10.9 \pm      7.8$ & $     4.7 \pm      7.4$ & $     0.3 \pm      3.2$ & $    -0.2 \pm      0.9$ \\
SMG15 & 14:23:40.95 +24:05:08.7 & $     2.4 \pm      1.3$ & $     3.1 \pm      3.3$ & $     6.6 \pm      9.1$ & $     6.8 \pm      8.2$ & $     8.2 \pm      7.4$ & $    -1.2 \pm      3.5$ & $     0.0 \pm      0.9$ \\
SMG16$^*$  & 14:23:53.69 +24:04:12.6 & $     6.2 \pm      1.4$ & $     7.3 \pm      3.4$ & $    18.0 \pm     10.5$ & $    -4.0 \pm      7.7$ & $    -3.5 \pm      7.5$ & $     3.4 \pm      3.3$ & $     0.6 \pm      0.9$ \\
SMG17$^*$  & 14:23:51.72 +24:05:48.8 & $     4.7 \pm      1.3$ & $     4.3 \pm      3.4$ & $   -14.6 \pm      9.7$ & $    -6.2 \pm      7.7$ & $    -7.4 \pm      7.4$ & $    -2.2 \pm      3.4$ & $    -0.1 \pm      0.9$ \\
\hline
\end{tabular}
\end{center}
{\small {\bf Notes.} $^*$ Sources for which the position is estimated based on the 100 $\mu$m PACS channel. $^{**}$ Fluxes which are not available due to the tSZ contamination.}
\label{tab:IR_ps}
\end{table*}

\subsection{Sub-millimeter point sources}\label{sec:smmps}
We make use of the Herschel Spectral and Photometric Imaging Receiver \citep[SPIRE;][]{griffin2010} and Photoconductor Array Camera and Spectrometer \citep[PACS;][]{poglitsch2010} data obtained during the Herschel Lensing Survey \citep[HLS,][]{egami2010,rawle2012}\footnote{Obs-IDs 1342188159, 1342188215 and 1342188216.} to identify sub-millimeter point sources and compute their expected spectral energy distribution (SED) as seen by NIKA, as described below. The Herschel data complement those from  NIKA, both in terms of wavelength 500, 350, 250, 160, and 100 $\mu$m, and angular resolution, FWHM = 35.2, 23.9, 17.6, 9.9, 6.1 arcsec, respectively.

The PACS maps were produced using the maximum likelihood map maker MADmap \citep{cantalupo2010}, provided as a PACS Photometer Level 2.5 Product. We used the Herschel Source List Product, generated with the software SUSSEXtractor \citep{savage2007}, which contains the location and flux of the sources found around \mbox{MACS~J1424.8+2404}. The sources are extracted independently for each frequency band, with a signal-to-noise threshold of 5. The 250 $\mu$m channel is the most complete with 15 sources detected in the $4 \times 4$ arcmin$^2$ around the cluster, and we used this 250 $\mu$m channel as a baseline to define the source positions and labels. The corresponding sources in the other channels were matched to the 250 $\mu$m ones on the basis of their positions. Two sources peaking at high frequency were not present in the 250 $\mu$m catalog and we relied on the 100 $\mu$m channel for their properties. By using all the Herschel frequency bands, we therefore found a total of 17 sub-millimeter sources, two of which correspond to the excesses seen in the NIKA 260~GHz map. Because of the relatively low resolution at 500 and 350 $\mu$m, a few sources are confused with their neighbors and, in general, not all the sources are identifiable in all the frequency bands in the catalog. To obtain the fluxes of all sources in all bands, we fit the amplitude of a Gaussian function to each map with the position fixed at the source reference location and the FWHM fixed to that of the respective Herschel channels. A local background was also fit. In order to account for the confusion in the flux uncertainties, we fit the same Gaussian function at random positions, where the noise is homogeneous, and use the dispersion as the uncertainty. We checked that the sources present in the catalog have compatible fluxes with respect to those we recovered. The positions and fluxes for all the sources are listed in Table~\ref{tab:IR_ps}.

Similarly, the fluxes of the sub-millimeter sources in the NIKA field were obtained by fitting the amplitude of a Gaussian model, using the NIKA FWHM at the source positions expected from Herschel data. These were then corrected for the filtering induced by the data reduction ($\sim15$ percent for point sources). Uncertainties were obtained from the standard deviation of the amplitudes recovered using Gaussian fits performed on the Monte-Carlo noise map realizations. The fluxes obtained and their uncertainties are summarized together with the Herschel data in columns of Table~\ref{tab:IR_ps}. By stacking the flux of all the sources, assuming that they are independent, we obtained an average flux of $1.96 \pm 0.82$ mJy at 260~GHz (1.15~mm) and $0.46 \pm 0.25$ mJy at 150~GHz (2.05~mm), corresponding to a mean detection of 2.4 and 1.9 $ \ \sigma$, respectively. If we exclude the two sources directly detected at the map level, the detection reduces to $1.59 \pm 0.89$ mJy at 260~GHz (1.8 $\sigma$) and $0.45 \pm 0.25$ mJy (1.8 $\sigma$) at 150~GHz.

\begin{table*}[h]
\caption{\footnotesize Offset with respect to the X-ray center, and flux extrapolated to the NIKA channels for the 17 sub-millimeter sources identified in the $4 \times 4$ arcmin$^2$ field around \mbox{MACS~J1423.8+2404}, obtained from the SED model described in Sect.~\ref{sec:smmps}. Despite the large uncertainties, these values allow to us estimate the impact of the contaminating sources on the reconstructed cluster properties.}
\begin{center}
\begin{tabular}{ccccccc}
\hline
\hline
Source &  R.A. offset & Dec. offset & 1.15 mm / 260~GHz & 2.05 mm / 150~GHz\\
 &  [arcsec] & [arcsec] & [mJy] & [mJy] \\
\hline
SMG01 &    -61.8 &     24.9 & $    1.52 \pm     0.64$ & $    0.30 \pm     0.19$ \\
SMG02 &     -5.0 &    -20.0 & $    2.39 \pm     1.10$ & $    0.62 \pm     0.42$ \\
SMG03 &    -78.1 &     85.1 & $    1.51 \pm     0.79$ & $    0.35 \pm     0.26$ \\
SMG04 &     73.7 &     -1.2 & $    1.05 \pm     0.67$ & $    0.25 \pm     0.21$ \\
SMG05 &      3.0 &      8.7 & $    2.01 \pm     0.97$ & $    0.50 \pm     0.34$ \\
SMG06 &    -75.5 &    -51.5 & $    1.13 \pm     1.12$ & $    0.31 \pm     0.39$ \\
SMG07 &     37.8 &     68.9 & $    0.60 \pm     0.57$ & $    0.15 \pm     0.17$ \\
SMG08 &    -18.6 &   -113.9 & $    2.53 \pm     1.17$ & $    0.60 \pm     0.40$ \\
SMG09 &     62.1 &   -109.8 & $    2.15 \pm     1.51$ & $    0.55 \pm     0.47$ \\
SMG10 &     44.5 &    -81.6 & $    3.79 \pm     1.56$ & $    0.97 \pm     0.55$ \\
SMG11 &    -86.8 &     52.2 & $    0.09 \pm     0.12$ & $    0.02 \pm     0.03$ \\
SMG12 &     60.1 &    -49.4 & $    0.05 \pm     0.07$ & $    0.01 \pm     0.02$ \\
SMG13 &    -31.9 &     97.4 & $    0.05 \pm     0.06$ & $    0.01 \pm     0.01$ \\
SMG14 &      6.1 &     69.6 & $    0.14 \pm     0.23$ & $    0.03 \pm     0.06$ \\
SMG15 &     93.8 &     28.7 & $    0.13 \pm     0.28$ & $    0.03 \pm     0.07$ \\
SMG16$^*$ &    -80.6 &    -27.4 & $    0.08 \pm     0.11$ & $    0.02 \pm     0.03$ \\
SMG17$^*$ &    -53.6 &     68.8 & $    0.03 \pm     0.04$ & $    0.01 \pm     0.01$ \\
\hline
\end{tabular}
\end{center}
{\small {\bf Notes.} $^*$ Sources for which the position is estimated based on the 100 $\mu$m PACS channel.}
\label{tab:IR_ps2}
\end{table*}

Only SMG02 and SMG06 are directly detected in the NIKA maps. To better constrain the fluxes in the NIKA bands, we modeled the SED with a modified blackbody spectrum, $F_{\nu} = A_0 \left(\frac{\nu}{\nu_0}\right)^{\beta_{\rm dust}} B_{\nu}(T_{\rm dust})$, where $A_0$ is a normalization, $\nu_0$ a reference frequency, $\beta_{\rm dust}$ the dust spectral index, and $T_{\rm dust}$ the dust temperature. We noticed a flux excess at 100 $\mu$m for most of the sources with respect to the best-fit SED derived from all the other channels, which we attribute to the modified blackbody spectrum being too simplistic a description of the data at high frequency. We therefore excluded this frequency for constraining the SED in the following. Since we aim to subtract the sources at 150~GHz, we also excluded the NIKA measurement at this frequency and only checked that predicted and measured fluxes are consistent. For the SMG02 and SMG05 sources, the 150~GHz data were not extracted because of strong contamination by the local tSZ signal. At larger radii, the faint, slowly varying tSZ signal is accounted for because we also fit for a local background. The tSZ impact on the fluxes of the other sources is therefore neglected and a bias is only expected in the case of significant tSZ signal at scales comparable to the beam.

We performed a Markov Chain Monte Carlo (MCMC) analysis to fit simultaneously for $\beta_{\rm dust}$ and $T_{\rm dust}$. This approach allows us to sample the corresponding parameter space and to automatically marginalize over $A_0$, which is highly degenerate with the two other parameters. The Metropolis-Hasting algorithm was used to sample the parameter space. Each model tested against the data was defined by a value of $\beta_{\rm dust}$ and $T_{\rm dust}$ and the normalization was linearly fit to the data. For NIKA, color corrections are expected to be negligible with respect to the calibration and statistical uncertainties ($\sim 1-2$\%). In the case of Herschel, we apply the color corrections given in \cite{poglitsch2010} for PACS and those available in the online documentation for SPIRE\footnote{\url{http://herschel.esac.esa.int/Docs/SPIRE/html/spire_om.html\#x1-830005.2.6}}. These coefficients were interpolated for each model with the values of $\beta_{\rm dust}$ and $T_{\rm dust}$, and provide a small correction to the Herschel fluxes ($\sim 5-10$\%). We finally obtained a set of models distributed with respect to the posterior likelihood. Each of the model SED spectra was integrated within the NIKA bandpasses with the distribution of fluxes giving the expected value and uncertainty. The results are summarized in Table~\ref{tab:IR_ps2}. A similar fitting procedure was developed by \cite{sayers2013c} with SPIRE data  to remove a contaminating signal from unresolved sources in Bolocam data. Our analysis also includes the PACS 160$\mu$m and  NIKA 260~GHz photometric values, which allows us to release the constraint on the slope parameter $\beta$ in contrast to the baseline value of 1.7 used by \cite{sayers2013c}.

\subsection{Surface brightness profile}\label{sec:surface_brightness_profiles_comparison}
\begin{figure*}[h]
\centering
\includegraphics[height=6.4cm]{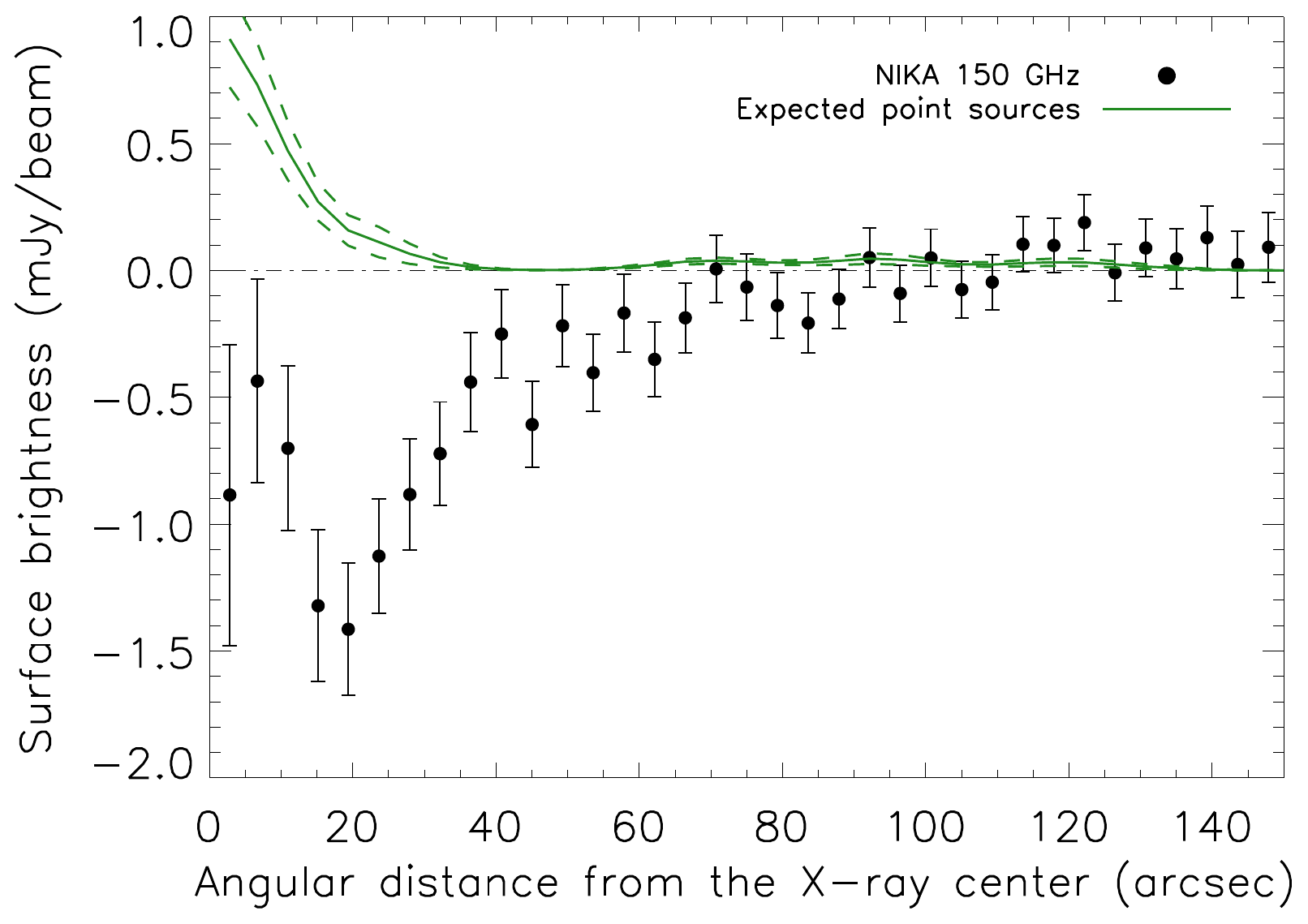}
\includegraphics[height=6.4cm]{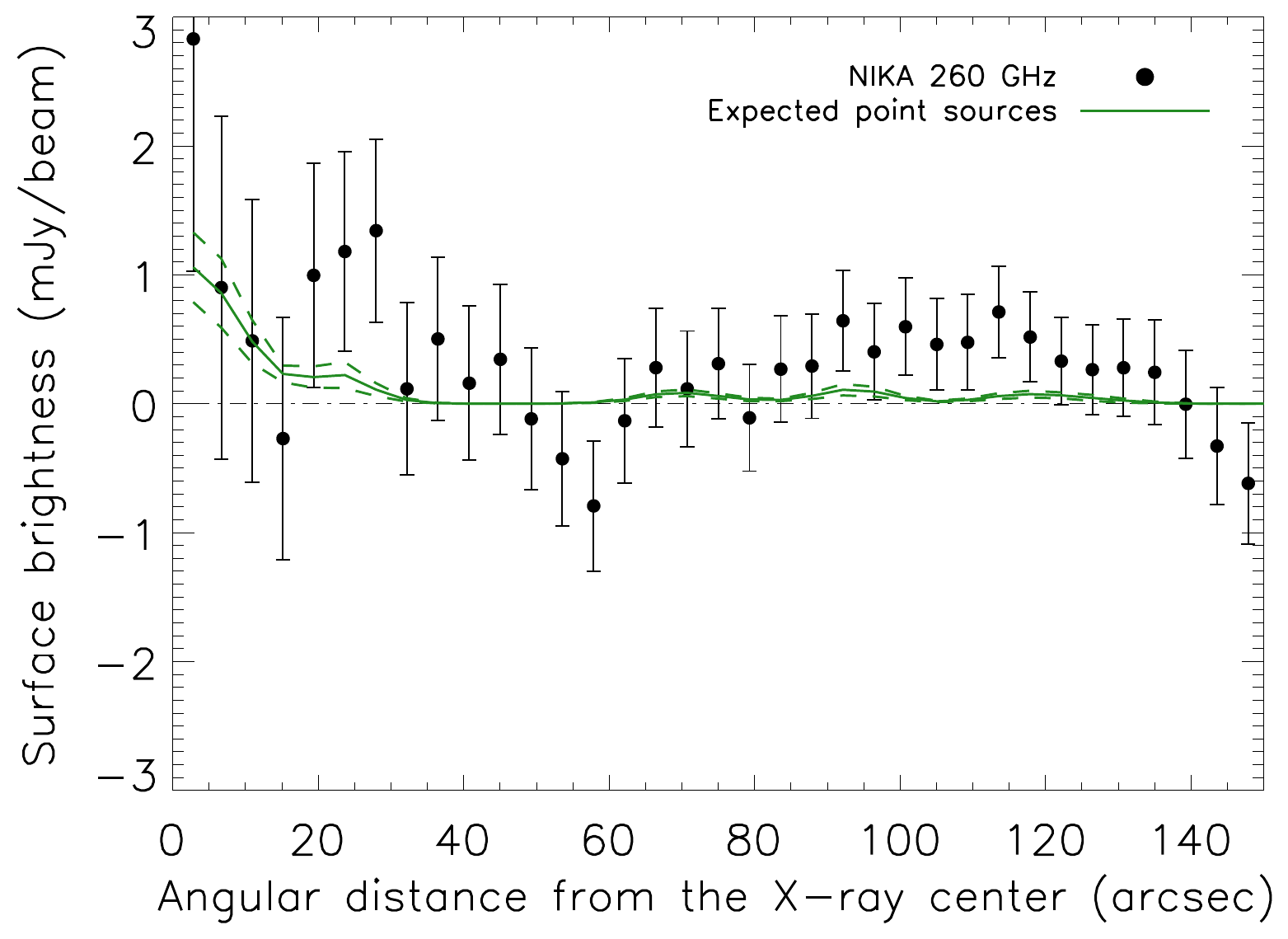}
\caption{\footnotesize Profiles at 150~GHz (left) and 260~GHz (right), in units of surface brightness, of the raw map (black dots) and the contribution expected from radio and sub-millimeter point sources (green solid line). The green dashed envelope gives the 68\% confidence interval on the point source profile.}
\label{fig:flux_profiles}
\end{figure*}

Figure~\ref{fig:flux_profiles}  shows the flux density profiles at 150 and 260~GHz, which is obtained by averaging the signal from the NIKA maps in Fig.~\ref{fig:flux_map} in radial bins around the X-ray center. Uncertainties were computed  using the noise realizations described in Sect.~\ref{sec:Raw_NIKA_observations}. The flux density profile was computed for each noise realization and the standard deviation of all the realizations per radial bin provided the associated error bars. This allows us to account for pixel--pixel noise correlation. We also computed the profile expected from the point sources for each NIKA band by simulating the corresponding maps using the point source fluxes and positions given in Tables~\ref{tab:Radio_ps2} and~\ref{tab:IR_ps2}. The profile was then calculated as for the NIKA maps and the error given by the dispersion of Monte Carlo realizations of the point source maps by randomly varying their fluxes within the Gaussian errors.

The 150~GHz profile decreases smoothly toward the center as a result of the tSZ signal, except in the inner 15 arcsec, where it rises. This is consistent with the positive signal expected from the presence of radio and sub-millimeter emission from point-like sources in the cluster core. Outside the core, the cluster is detected, at the profile level, up to 60 arcsec radius. 

The sub-millimeter sources shown in Fig.~\ref{fig:MACSJ1424_mutiw} are located in two distinct regions. Two of them are within 30 arcsec from the X-ray peak, where we expect the tSZ contribution to be the strongest, while the others are concentrated in a ring of 70 arcsec inner radius and 130 arcsec outer radius. No source is seen around 50 arcsec from the cluster center. This source distribution is reproduced, despite the large uncertainties in the 260~GHz profile. There is an excess near the center, which we attribute to the sum of tSZ and sub-millimeter sources (one of  which is detected with NIKA at the map level). A dip can be  seen around 1 arcmin, and another excess is seen around 100 arcsec from the center, both due to the distribution of Herschel point sources. Overall, the NIKA 260~GHz profile is consistent with that expected from the observed point source distribution.  A deficit of sub-millimeter surface brightness was also observed toward four galaxy clusters by \cite{zemcov2013} after the removal of the detected sources. This phenomenon is due to the gravitational lensing of the cosmic infrared background emission induced by the clusters, and might contribute to the distribution we observe toward \mbox{MACS~J1423.8+2404}.

\section{XMM-Newton and Chandra X-ray data reduction}\label{sec:XMM_Newton_and_Chandra_X_ray_data_reduction}
X-ray observations of galaxy clusters are sensitive to both the electronic density and the temperature of the ICM. The X-ray surface brightness is expressed as 
\begin{equation}
        S_X = \frac{1}{4 \pi \left(1+z\right)^4} \int n_e^2 \Lambda(T_e, Z) \ dl,
        \label{eq:sx}
\end{equation}
where $\Lambda(T_e,Z)$ is the cooling function, which depends on the temperature and the ICM metallicity $Z$, and is roughly proportional to $T_e^{1/2}$. While X-ray imaging is mainly sensitive to the electronic density, the gas temperature can be estimated from X-ray spectroscopy. See for example \cite{bohringer2010} for a review.

\subsection{Data preparation}
\mbox{MACS~J1423.8+2404} was observed by the Chandra Advanced CCD Imaging Spectrometer (ACIS) for 115 ks (obs-ID $4195$) and by the XMM-Newton European Photon Imaging Camera (EPIC) for 109 ks in total (obs-IDs $720700301$ and $720700401$). XMM-Newton datasets were processed by applying the latest calibration files via the Science Analysis System (SAS) version $14.0.0$ and the calibration files available in May $2015$. Chandra datasets were processed with the Chandra Interactive Analysis of Observation (CIAO) software suite version $4.7$ and calibration database version $4.6.5$.  We applied the very faint mode\footnote{\label{fn1}\url{http://cxc.cfa.harvard.edu/cal/Acis/Cal\_prods/vfbkgrnd/}} filtering to the Chandra datasets to reduce contamination from the stationary flux of high
energetic particles.  We rejected events for which the keyword {\sc pattern}  is $> 4$ (for MOS$1,2$) and $>13$ (for PN) for XMM-Newton. 

The Chandra and XMM-Newton datasets were analyzed using the same technique, so unless otherwise stated, the procedures described in the following were applied to both datasets. To remove solar soft proton flare contamination, we followed the procedures described in \cite{pratt2007} and in the Chandra {\sc cookbook}\footnote{\url{http://cxc.harvard.edu/contrib/maxim/acisbg/COOKBOOK}} for the XMM-Newton and Chandra datasets, respectively, rejecting time intervals where the count rate exceeded $3 \ \sigma$ with respect to the mean value. We found no flare contamination in the Chandra dataset, thus, the full 115~ks observation was used; for the XMM-Newton dataset, the useful exposure time was 122 ks for MOS$1$+MOS$2$ and 41 ks for PN. We ran a wavelet detection algorithm with a threshold of $5 \ \sigma$ on exposure corrected images in the $[0.7-7]$ keV and $[0.3-2]$ keV bands Chandra and XMM-Newton, respectively, to identify point sources. The resulting source lists were merged, inspected by eye, and used as a mask to remove point source contributions from the analysis.

\subsection{Vignetting correction}
To correct for vignetting, we followed the procedure described in \cite{arnaud2001}, computing a weight for each photon, defined as the ratio of the effective area at the aimpoint to that at the photon position. Using these quantities, we can perform imaging and spectroscopic analysis as if the detector had a flat response equal to that at the aimpoint.  The weights were computed using the \verb?evigweight? routine for XMM-Newton;  we used the procedure described in \textcolor{blue}{Bartalucci et al. (in prep.) }for Chandra.  We also computed the effective exposure time for each photon, producing exposure maps that take bad pixels and columns into account . The response files for weights and spectroscopic analysis were computed with the appropriate tools in CIAO (\verb?mkarf? and \verb?mkacisrmf?) and SAS (\verb?arfgen? and \verb?rmfgen?).

\subsection{Background subtraction}
The X-ray background is due to diffuse sky emission and an instrumental component caused by the interaction of high energetic particles with telescope instruments. To estimate the instrumental component, we used datasets tailored to isolate this component, namely {\sc closed} and {\sc stowed} for XMM-Newton and Chandra, respectively. To match the observation, these datasets were  skycasted and normalized in the $[10-12], \ [12-14]$ keV and in the $[9.5-10.6]$ keV band for MOS$1,2$-PN and ACIS, respectively.  We used background datasets from period D for Chandra, since the observation was performed in 2003. We computed the weights for the instrumental background datasets by applying the same point source masking and filtering procedures as for the observation datasets. We then subtracted from the surface brightness and temperature profiles the instrumental background evaluated using these datasets.

The residual background component is composed of thermal Galactic emission \citep{snowden1995} and the blending of unresolved distant point sources, known as the cosmic X-ray background \citep{giacconi2001}.  We determined a region free from cluster emission and subtracted the residual mean background count rate for the surface
brightness profile analysis. For the spectroscopic analysis, we extracted the spectrum from the source-free region and fitted it with a multicomponent model composed of two absorbed {\sc MEKAL} thermal components and an absorbed power law (for details on the model used, see \citealt{pratt2009}). When performing further spectroscopic analysis we added the background model as a fixed component with an amplitude scaled by the ratio of the area of the region of interest to that of the source-free background area. The X-ray spectra were extracted and analyzed in the $[0.3-10]$ keV and $[0.7-10]$ keV bands for XMM-Newton and Chandra, respectively. 

\section{Multiwavelength comparison}\label{sec:Multi_wavelength_comparison}
\begin{figure*}[h]
\centering
\includegraphics[trim=0cm 0cm 0cm 0cm, clip=true, width=0.99\textwidth]{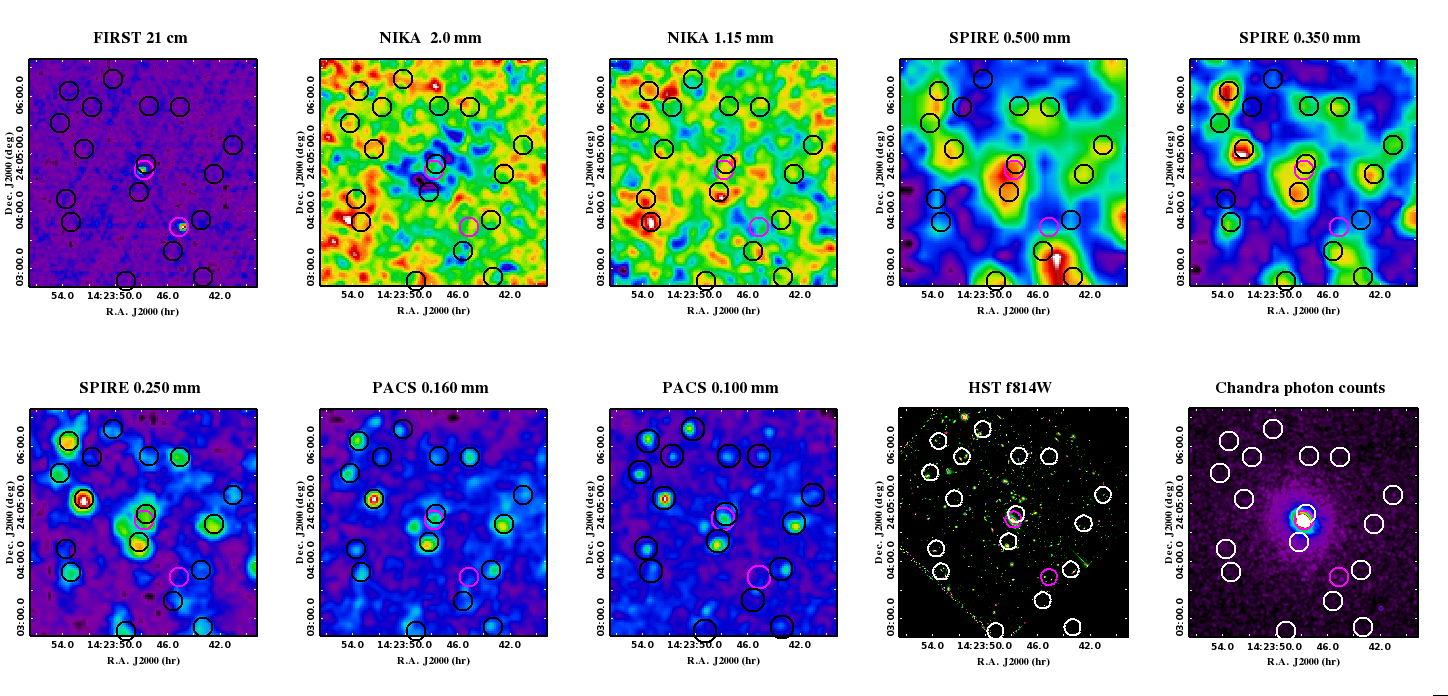}
\caption{\footnotesize Multiwavelength dataset of \mbox{MACS~J1423.8+2404}. The origin of the data is given on top of each map. The maps have been smoothed and their range adapted for visualization purposes. The displayed PACS and SPIRE maps have been reprojected to match the NIKA map grids. The 10 arcsec radius circles show the point source locations in black/white for sub-millimeter sources (table~\ref{tab:IR_ps}) and magenta for radio sources (table~\ref{tab:Radio_ps}).}
\label{fig:MACSJ1424_mutiw2}
\end{figure*}

Fig.~\ref{fig:MACSJ1424_mutiw2} presents a multiwavelength overview of \mbox{MACS~J1423.8+2404}, including the {\sc first} survey \citep[Faint Images of the Radio Sky at Twenty-Centimeters;][]{becker1995} observations at 1.4~GHz that provided the radio point source locations. The two radio sources discussed in Sect.~\ref{sec:smmps} are clearly visible on the map. The complementarity of the Herschel and NIKA instruments is clear; the NIKA bands complement the Herschel spectral coverage at lower frequencies. In addition, the angular resolution of the NIKA bands is comparable to that of SPIRE at 250 $\mu$m and PACS at 160 $\mu$m, allowing us to break the confusion in the low frequency SPIRE bands in the crowded cluster environment for sources that are directly detected by NIKA. The Hubble Space Telescope (HST) data from the Cluster Lensing And Supernova survey with Hubble program \citep[CLASH;][]{postman2012} show the galaxy distribution. The (smoothed) Chandra X-ray image traces the gas electronic density and is used further in Sect.~\ref{sec:Radial_pressure_reconstruction}.

\begin{figure}
\centering
\includegraphics[trim=1cm 0cm 5cm 2cm, clip=true, width=0.45\textwidth]{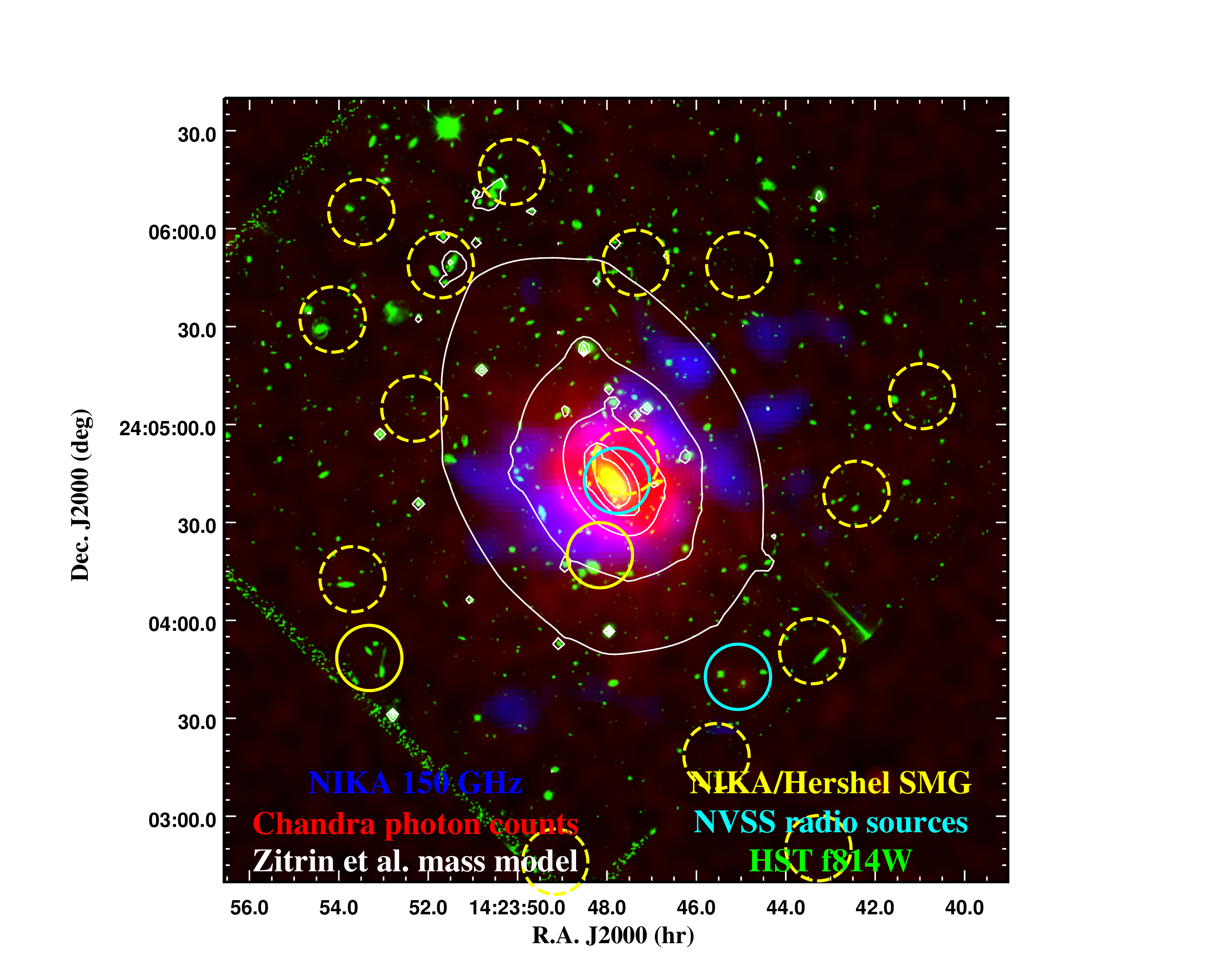}
\caption{\footnotesize Composite multiwavelength overview image of \mbox{MACS~J1423.8+2404}. {\bf Blue image}: NIKA 150~GHz map showing the tSZ signal. {\bf Red image}: Chandra photon counts (Obs-ID 04195) tracing the electronic density. {\bf White contours}: Surface mass distribution model obtained by \cite{zitrin2011,zitrin2015} on a linear scale. {\bf Yellow circles}: (Sub-)millimeter sources candidate locations obtained using the NIKA 260~GHz map (solid line) and identified using Herschel (dashed-line). {\bf Cyan circle}: Location of the radio point sources present in the field as obtained from VLA \citep{laroque2003}. {\bf Green image}: Hubble Space Telescope image using the F814W filter obtained by the CLASH  program \citep{postman2012} showing the location of the galaxies.}
\label{fig:MACSJ1424_mutiw}
\end{figure}

Fig.~\ref{fig:MACSJ1424_mutiw} shows a composite multiwavelength image of the cluster. The figure includes the NIKA 150~GHz tSZ map, the Chandra X-ray photon counts, the NIKA and Herschel sub-millimeter galaxy locations, the NVSS radio source locations, and the HST image from CLASH. We also provide, for visual comparison, the surface mass distribution model of \mbox{MACS~J1423.8+2404} produced by \cite{zitrin2011,zitrin2015} with the CLASH data. The image provides a detailed picture of the cluster complementing the discussion of Sect.~\ref{sec:Introduction}. The NIKA 150~GHz  tSZ signal surrounds the cluster core. The hole seen in the tSZ signal is coincident with the BCG and results from  canceling  the tSZ by the radio and sub-millimeter signal, as discussed in Sect.~\ref{Radio_and_infrared_point_sources}. The X-ray morphology is very peaked and the maximum of the emission coincides with the BCG. The ellipticity of the spatial mass distribution is very clear from the strong lensing map, and is also visible in the X-ray map but less significant.\ The ellipticity is not visible in the tSZ map because of the limited signal-to-noise and the point source contamination. The galaxy distribution does not show any particular group that would be the sign of a merging event.

\section{Radial thermodynamical reconstruction}\label{sec:Radial_pressure_reconstruction}
The radial physical properties of the ICM were reconstructed using two approaches. The first, detailed in Sect.~\ref{sec:X_ray_extraction_of_the_cluster_radial_thermodynamic_profiles},  uses the X-ray data alone. The electronic density and  gas temperature are directly measured and serve to reconstruct all the other quantities. This approach strongly depends on the spectroscopic temperature reconstruction. The second approach, detailed in Sect.~\ref{sec:modeling}, consists of  jointly fitting the tSZ data and the deprojected electronic density extracted from the X-ray data. The primary quantities are therefore the pressure and electronic density, and as such the method depends only weakly on X-ray spectroscopy through  electronic density reconstruction. The comparison of the two approaches is given in Sect.~\ref{sec:results}.

\subsection{X-ray radial thermodynamic profiles}\label{sec:X_ray_extraction_of_the_cluster_radial_thermodynamic_profiles}
We need to determine the temperature and electronic density profiles to compute the X-ray electronic pressure and entropy profiles. The electronic density profiles was computed by applying the regularized deconvolution and deprojection technique described in \cite{croston2006}. We extracted the surface brightness profile from concentric annuli centered on the X-ray peak, (R.A., Dec.) = (14:23:47.9, +24:04:42.3), in the $[0.3-2.5]$ keV and $[0.7-2.5]$ keV bands from the three combined EPIC and ACIS cameras, respectively. After background subtraction, the profiles were rebinned via a logarithmic binning factor of $1.05$ to have a $3 \ \sigma$ significance for each bin, and were point spread function (PSF) deconvolved using the analytical model of \cite{ghizzardi2001} for XMM-Newton. For Chandra, we assumed that the PSF is negligible with respect to the width of the bins. 

The deconvolved, deprojected surface brightness profiles were converted to density using a conversion factor determined from the projected temperature profile. This temperature profile was determined by extracting spectra from concentric annuli centered on the X-ray peak, and the width of each annulus was defined to have a signal-to-noise ratio of $30$ after background subtraction. We measured the temperature in each bin by fitting the spectrum with an absorbed {\sc MEKAL} model, where the absorption was fixed to $N_{H} = 2.2 \times 10^{20} \ {\rm cm}^{-3}$ \citep{kalberla2005}, the redshift to $z=0.545$, adding the scaled sky background component discussed above.  We fit with the parametric model described in \cite{vikhlinin2006} to interpolate the temperature profile for each
surface brightness bin. The emissivity term, $\Lambda$ term in equation~\ref{eq:sx}, depends weakly on temperature, so that the change we obtain on the density profile using Chandra temperature profile instead of XMM is less than 1\% over the entire radial range. The differences in temperature reconstruction between Chandra and XMM-Newton \citep[e.g.,][]{sch15} thus do not affect the electronic density. To deproject the temperature profile, we measured the spectroscopic-like temperature \citep{mazzotta2004} using the weighting scheme implemented in \cite{vikh_multit}, where the observed temperature profile is modeled as the weighted sum of concentric plasma shells each at a different temperature. As for the density profiles, we take  the PSF effects for XMM-Newton into account. 

Finally, $M_{500}$, the cluster mass enclosed within $R_{500}$\footnote{$R_{500}$ is the radius within which the mean cluster density is equal to 500 times the critical density of the Universe at the cluster's redshift.}, was calculated by iteration about the $M_{500}$--$Y_X$ relation of \cite{arnaud2010}.

\subsection{Modeling of the ICM and joint tSZ/X-ray fitting procedure}\label{sec:modeling}
To fit the tSZ and electronic density jointly, we modeled the ICM using the approach described in detail in \cite{adam2014}. As seen in Sect.~\ref{sec:Multi_wavelength_comparison} and as emphasized by \cite{morandi2010}, \mbox{MACS~J1423.8+2404} is elliptical. Nevertheless, we assume spherical symmetry because this work does not focus on its geometry, but on the impact of the presence of point sources on the pressure profile reconstruction. Moreover, the significance of the NIKA 150~GHz tSZ map is not sufficient to constrain any asymmetry. 

The radial distribution of the cluster electronic pressure was modeled by a gNFW profile \citep{nagai2007}, described by
\begin{equation}
        P_e(r) = \frac{P_0}{\left(\frac{r}{r_p}\right)^c \left(1+\left(\frac{r}{r_p}\right)^a\right)^{\frac{b-c}{a}}}.
\label{eq:gNFW}
\end{equation}
The parameter $P_0$ is a normalization constant; $r_p$ is a characteristic radius; and $a$, $b$, and $c$ set the slopes at intermediate, large, and small radii, respectively. This model was chosen to allow the description of the profile at all scales. The electronic density was modeled by a \emph{Simplified Vikhlinin Model} \citep{vikhlinin2006},
\begin{equation}
        n_e(r) = n_{e0} \left[1+\left(\frac{r}{r_c}\right)^2 \right]^{-3 \beta /2} \left[ 1+\left(\frac{r}{r_s}\right)^{\gamma} \right]^{-\epsilon/2 \gamma}.
\label{eq:SVM}
\end{equation}
The parameter $n_{e0}$ is the central density, $r_c$ is the core radius, and $\beta$ is related to the slope of the profile. The second term allows a steepening of the profile at large scales. The parameter $\epsilon$ gives the change in the slope, $r_s$ the radius at which the transition occurs, and $\gamma$ the width of the transition. In the following, we set $\gamma = 3$ since this value provides a good fit to all clusters considered in the analysis of \cite{vikhlinin2006}. All the other parameters are varied when fitted to the data.

With the pressure and the density in hand, we compute the temperature profile assuming the ideal gas law, $k_B T_e(r) = \frac{P_e(r)}{n_e(r)}$, and the entropy profile as $K(r) =  \frac{P_e(r)}{n_e(r)^{5/3}}$. The total mass, assuming hydrostatic equilibrium, enclosed within $r$ is then given by 
\begin{equation}
M_{\rm HSE}(r) = -\frac{r^2}{\mu_{gas} m_p n_e(r) G} \frac{dP_e(r)}{dr}
,\end{equation}
where $m_p$ is the proton mass, $G$ is Newton‚Äôs constant, and $\mu_{gas} = 0.61$ the mean molecular weight of the gas.

The parameter space was sampled using the Markov Chain Monte Carlo (MCMC) approach detailed in \cite{adam2014}, jointly fitting the 150~GHz NIKA tSZ map and the electronic density profile computed from the X-ray data. We added an additional constraint on the total tSZ flux of the cluster with the Planck Compton parameter map \citep{planck2015XXII}. The tSZ models were convolved with the effective transfer function of the observations, including the beam smoothing and the large-scale filtering cutoff due to the removal of the atmospheric noise. \mbox{MACS~J1423.8+2404} is not present in the Planck catalogue of tSZ sources \citep{planck2015XXVII}, but we obtained an upper limit on its flux by integrating the Compton parameter map \citep{planck2015XXII} using aperture photometry. The error on the flux was computed by performing the same measurement randomly around the cluster, where the noise is homogeneous. The flux was measured to be $Y_{\rm tot}^{\rm Planck} = \left(0.40 \pm 0.66\right) \times 10^{-3}$ arcmin$^2$. The MCMC sampling procedure also marginalizes over nuisance parameters such as the zero level of the NIKA map, the calibration uncertainty, and the central point source flux and position when included in the fit. Full details can be found in \cite{adam2014}.

The constraints obtained on the pressure and electronic density profiles are almost independent. However, each model compared to the data includes relativistic corrections computed using the radial temperature profile, given by the ratio of the electronic pressure and density profiles, for each radial shell of the ICM. This  essentially affects the constraint on the pressure profile but the effect is very small compared to the uncertainties. Therefore, the constraint on the electronic density profile is driven by the X-ray data and that on the pressure is largely driven by the tSZ data. The Planck constraint on the overall tSZ flux is relatively weak because of the location of the cluster on the sky and the noise inhomogeneity. However, Planck provides an upper limit that allows the MCMC procedure to avoid models that diverge at large scales, where NIKA is not sensitive. Planck and NIKA are therefore highly complementary to constrain the pressure profile from small to large scales.

\subsection{Results}\label{sec:results}
\mbox{MACS~J1423.8+2404} is known to be a typical cool core \citep[e.g.,][]{morandi2010}, so we used the pressure profile parameters found for such clusters by \cite{arnaud2010} as a baseline: $\left(a,b,c\right) = \left(1.2223, 5.4905, 0.7736\right)$. The X-ray photon count provided by XMM-Newton is larger than that of Chandra, allowing us to probe the cluster ICM up to larger radii. We therefore used the XMM-Newton results as a reference and we cross-checked our results with the Chandra data.

\subsubsection{Impact of the point sources at millimeter wavelengths}
\begin{figure}[h]
\centering
\includegraphics[width=0.49\textwidth]{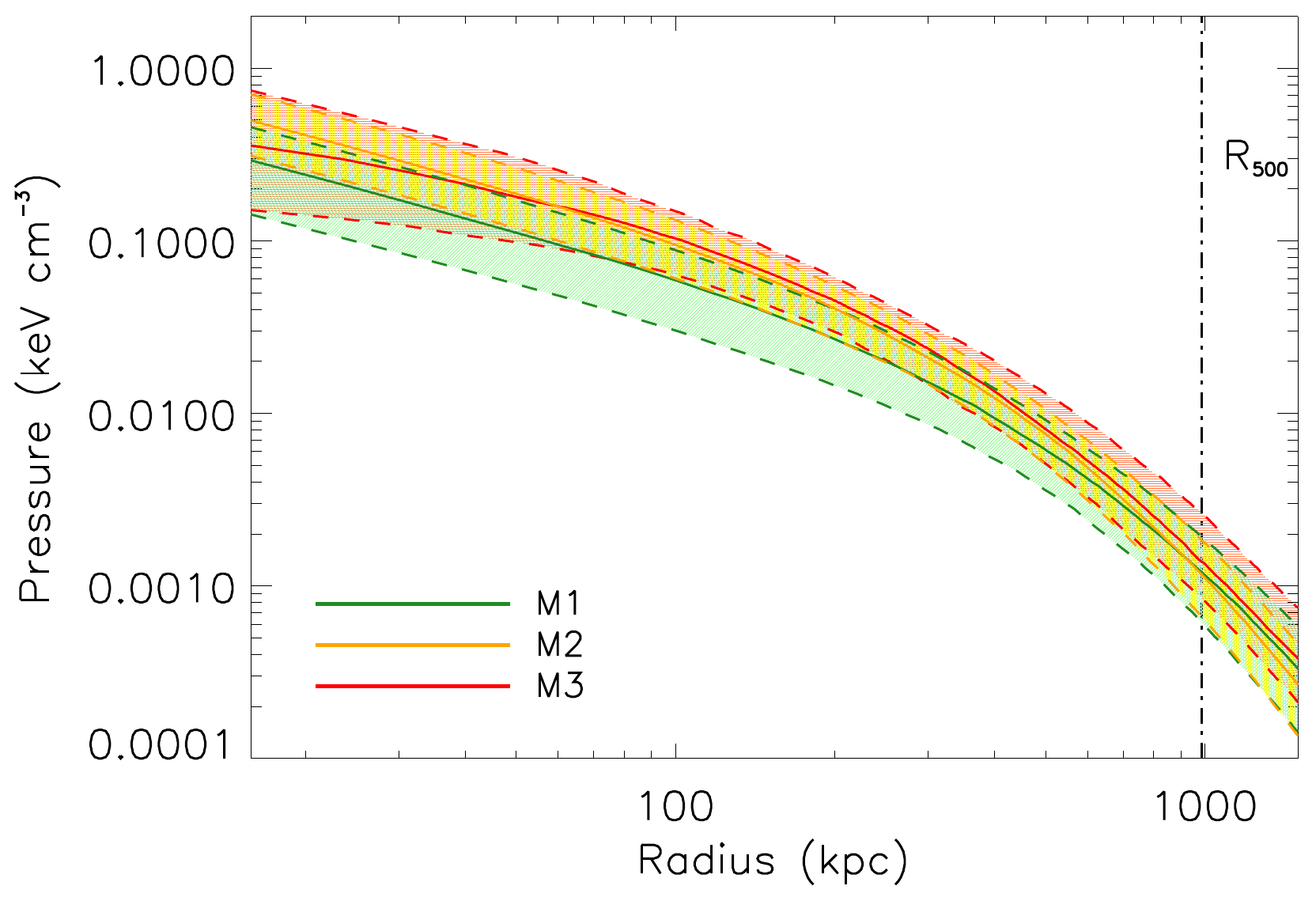}
\caption{\footnotesize Constraints on the pressure profile of \mbox{MACS~J1423.8+2404} for  three models M1 (green), M2 (orange), and M3 (red). The solid lines gives the central profiles and dashed line represent the 68\% confidence limits.}
\label{fig:MACSJ1424_pressure_point_source}
\end{figure}

The impact of the point source contamination on the reconstructed pressure profile was tested by considering three different cases: 
\begin{enumerate}
\item The presence of point sources was ignored and we fit the parameters $P_0$ and $r_p$, keeping the slope parameters fixed to their baseline values. This case is referred to as model 1 (M1).
\item The point sources were subtracted assuming the fluxes of Table \ref{tab:IR_ps2} and we repeated the fit of model M1. This case is referred to as model 2 (M2).
\item The point sources were subtracted assuming the fluxes given in Table \ref{tab:IR_ps2}, but possible residuals of the central sources were also fitted. In this case, we also released the constraints on the inner and outer slope parameters, $c$ and $b$, which are also fitted. The intermediate slope parameter $a$ was held fixed because it is strongly degenerate with the characteristic radius $r_p$ and the outer slope $b$. This case is referred to as model 3 (M3).
\end{enumerate}
Comparison of the output pressure profile from models M1 and M2 (see Fig.~\ref{fig:MACSJ1424_pressure_point_source}) allows a direct estimation of the impact of point sources. The use of model M3 was motivated by the large uncertainties on the point source fluxes as shown in Table \ref{tab:IR_ps2}. As discussed in Sect.~\ref{sec:surface_brightness_profiles_comparison} and shown on Fig.~\ref{fig:flux_profiles}, the sources that are in the outer region of the cluster do not contribute significantly to the radial flux density. However, this is not the case for the central sources, in particular, for RS1 and SMG05. We expect a significant correlation between the flux of these sources and the pressure profile parameters related to the cluster core, such as $P_0$, $r_p$, and $c$. Those related to the external regions of the profile are also affected because of projection effects. Thus model M3 allows us to test the constraints that can be obtained on the pressure profile in the cluster core in the presence of a contaminating central point source, when no strong prior on its flux is available, as is the case here.

Figure~\ref{fig:MACSJ1424_pressure_point_source} shows the resulting pressure profiles. Model M1 is lower than the two others at all scales except above 1000 kpc, where NIKA does not provide a direct constraint because of the large-scale filtering. This corresponds to the fact that point sources and, in particular those close to the core, compensate for the overall flux amplitude of the cluster. Apart from its normalization, the shape of the pressure profile is similar for M1 and M2, as the constraint on $r_p$ is not significantly affected by the point source subtraction. The effect of the source in the cluster core is then diluted over the entire  profile via the $P_0$ -- $r_p$ degeneracy. The results for M3 are close to those for M2, but the error contours are larger because of the extra freedom available in the parameter space. This is particularly true in the inner region, where the error on the flux of the fitted point source propagates into the profile via the parameter degeneracies. The inner pressure distribution thus remains poorly constrained unless the point source is subtracted with sufficient accuracy. At large scales, around 800 kpc,  M3 is slightly above M2. However, the differences observed between models are not significant with respect to the uncertainties, which are dominated by the low exposure time spent with NIKA on this cluster.

\begin{figure*}[h]
\centering
\includegraphics[width=0.49\textwidth]{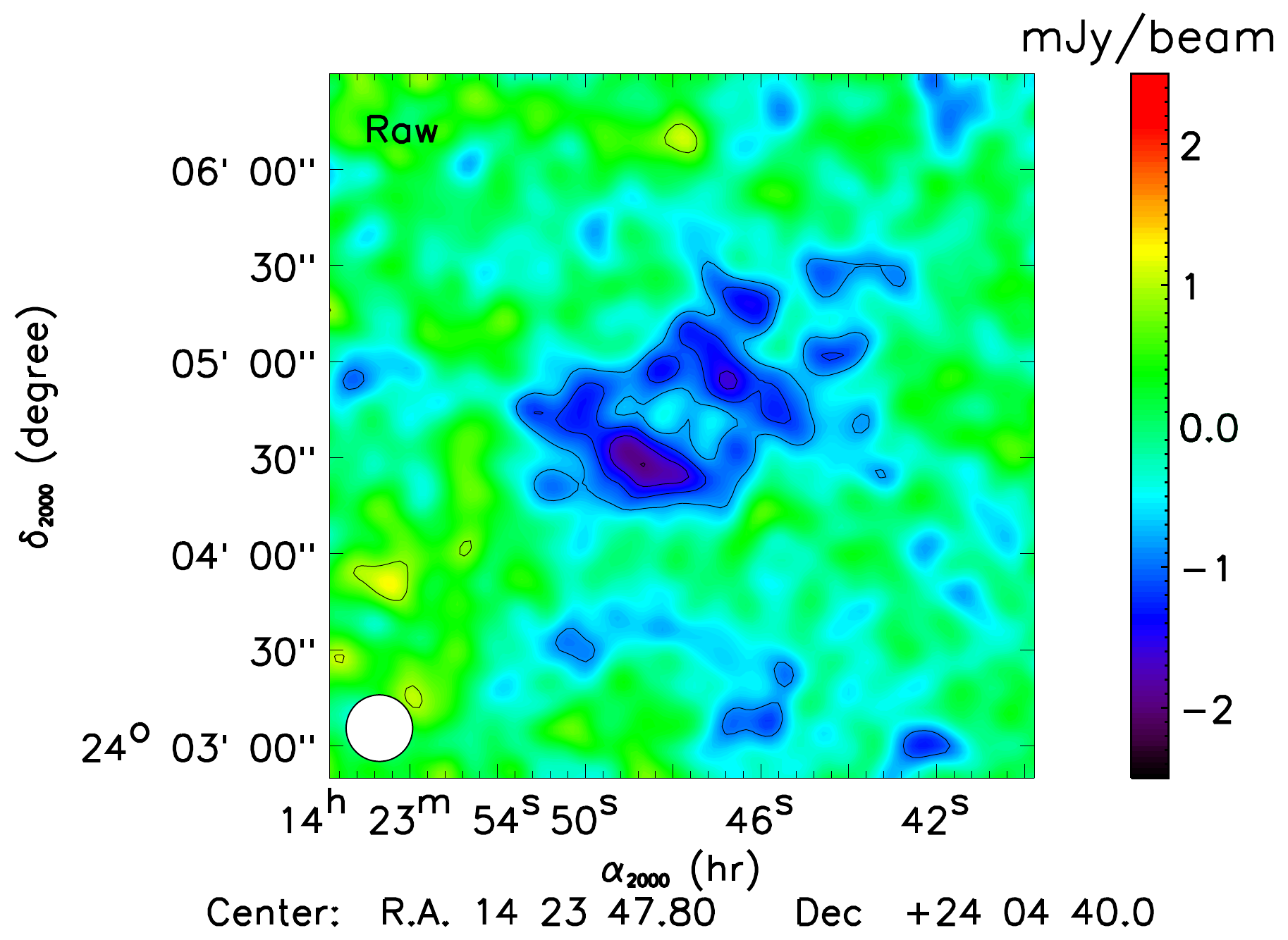}
\includegraphics[width=0.49\textwidth]{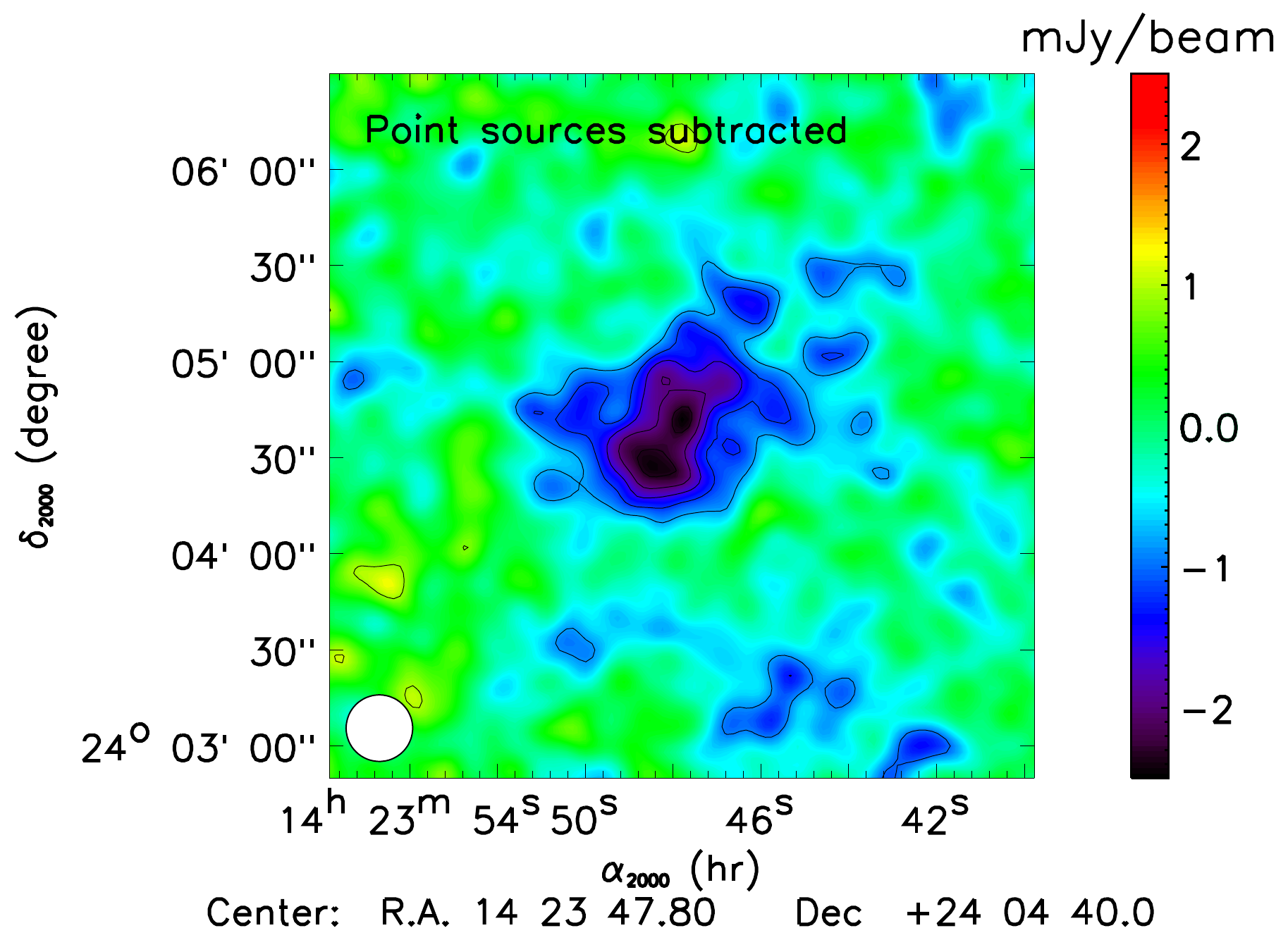}
\includegraphics[width=0.49\textwidth]{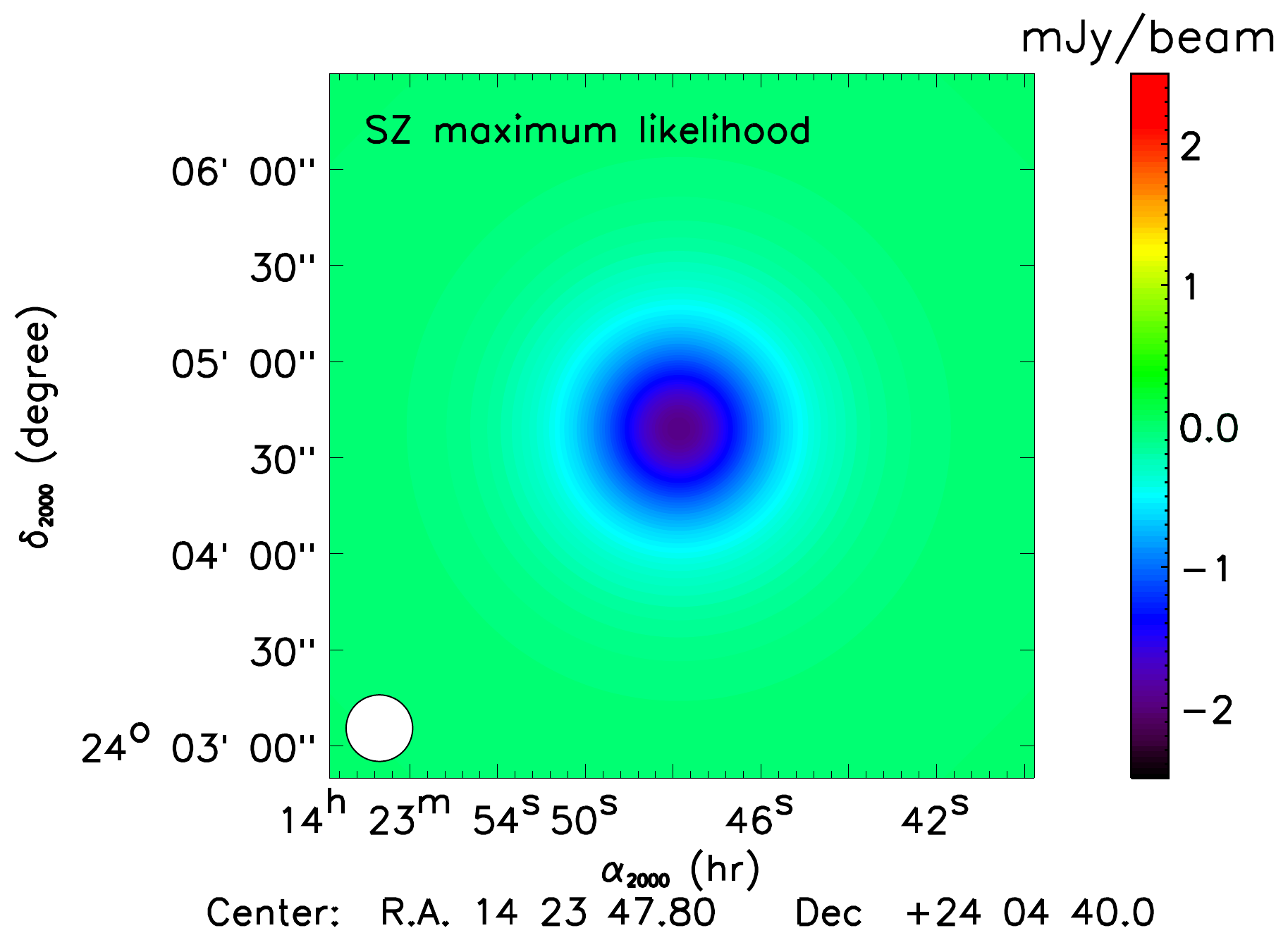}
\includegraphics[width=0.49\textwidth]{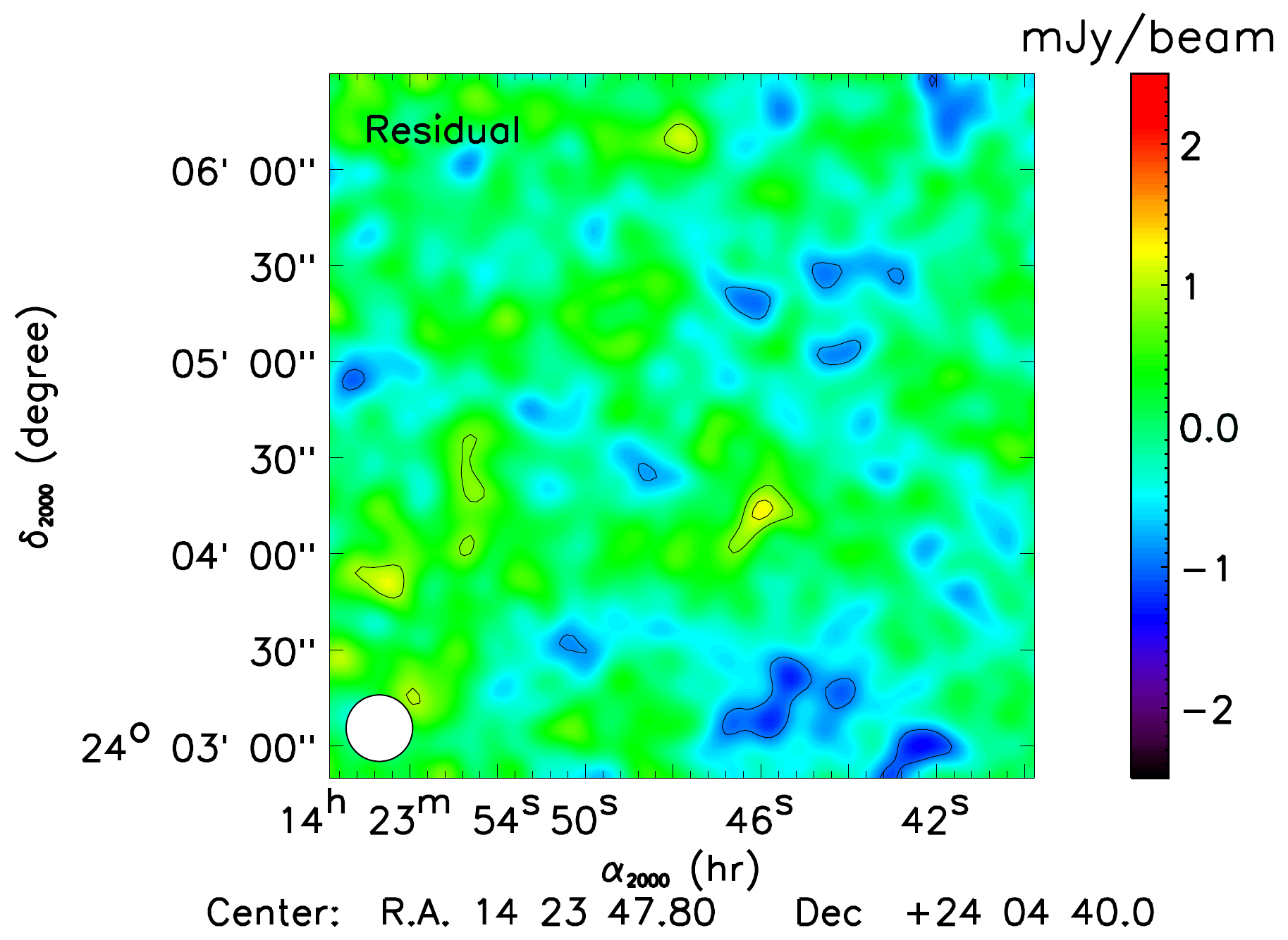}
\caption{\footnotesize Comparison of the raw 150~GHz input NIKA map (top left), which is the same map after point source subtraction (top right), the best-fit tSZ model (bottom left), and the final residual map. Contours are the same as in Fig.~\ref{fig:flux_map} and give the significance in unit of $\sigma$, starting from $\pm 2 \ \sigma$. }
\label{fig:MACSJ1424_MCMC_modeling}
\end{figure*}

To search for possible deviations from a spherical relaxed morphology, we subtracted the best-fit model of M3 from the NIKA 150~GHz map. Fig.~\ref{fig:MACSJ1424_MCMC_modeling} shows the resulting raw, point source subtracted, tSZ best-fit, and residual 150~GHz maps. After removing the point sources (fluxes in Table \ref{tab:IR_ps2}) and the best-fit of model M3, the cluster appears to be fairly compact. The small excess toward the south has a significance of less than 2 $\sigma$ and can be attributed to a poor subtraction of SMG05, whose position is very close on the map. The signal is overall relatively spherical, as confirmed by the residual map, which is consistent with the noise. Deeper tSZ observations would be necessary  to further constrain the morphology of the cluster, in particular in terms of ellipticity.

\subsubsection{Comparison between tSZ and X-ray derived cluster thermodynamics}
\begin{figure*}[h]
\centering
\includegraphics[height=6.3cm]{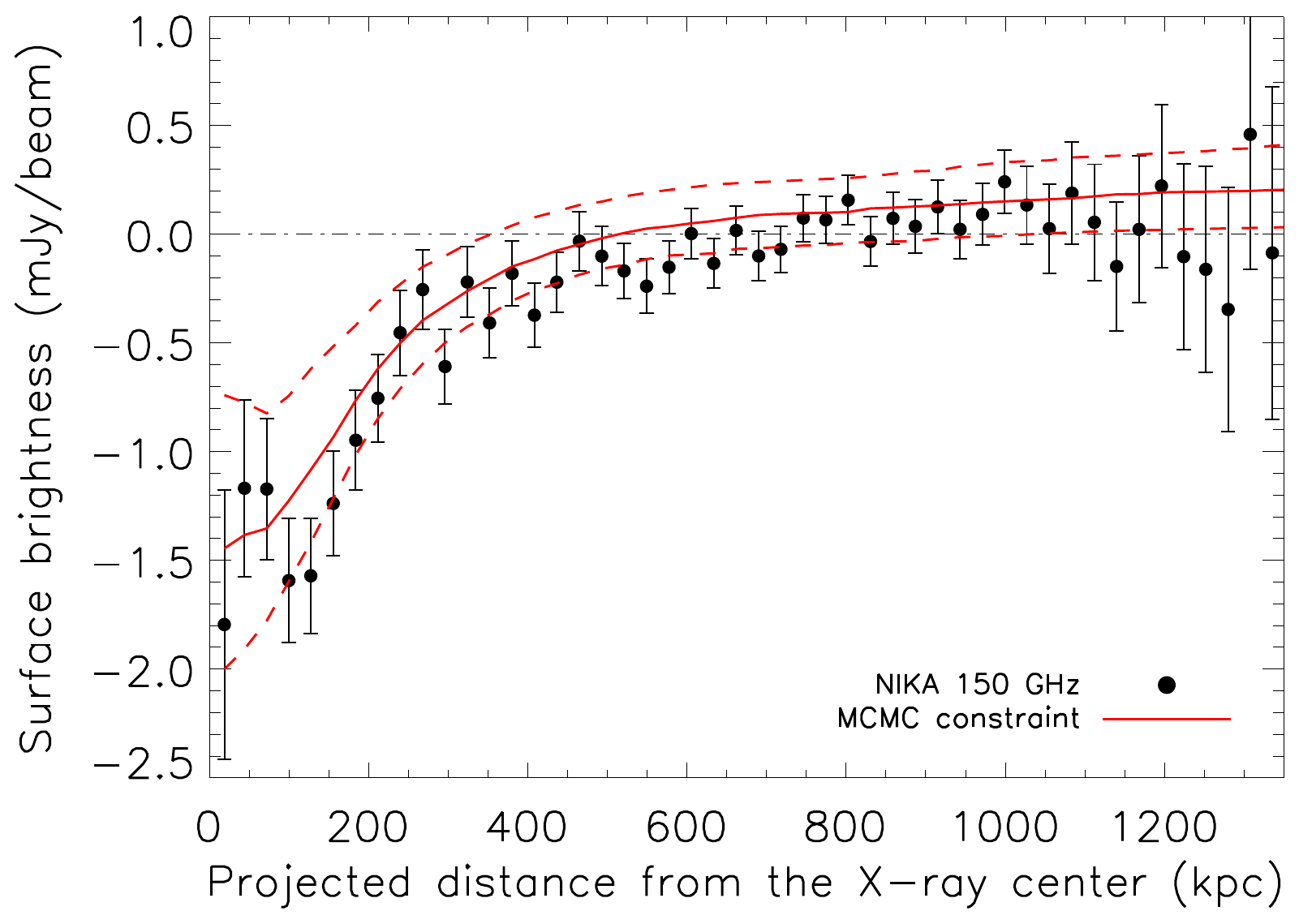}
\includegraphics[height=6.3cm]{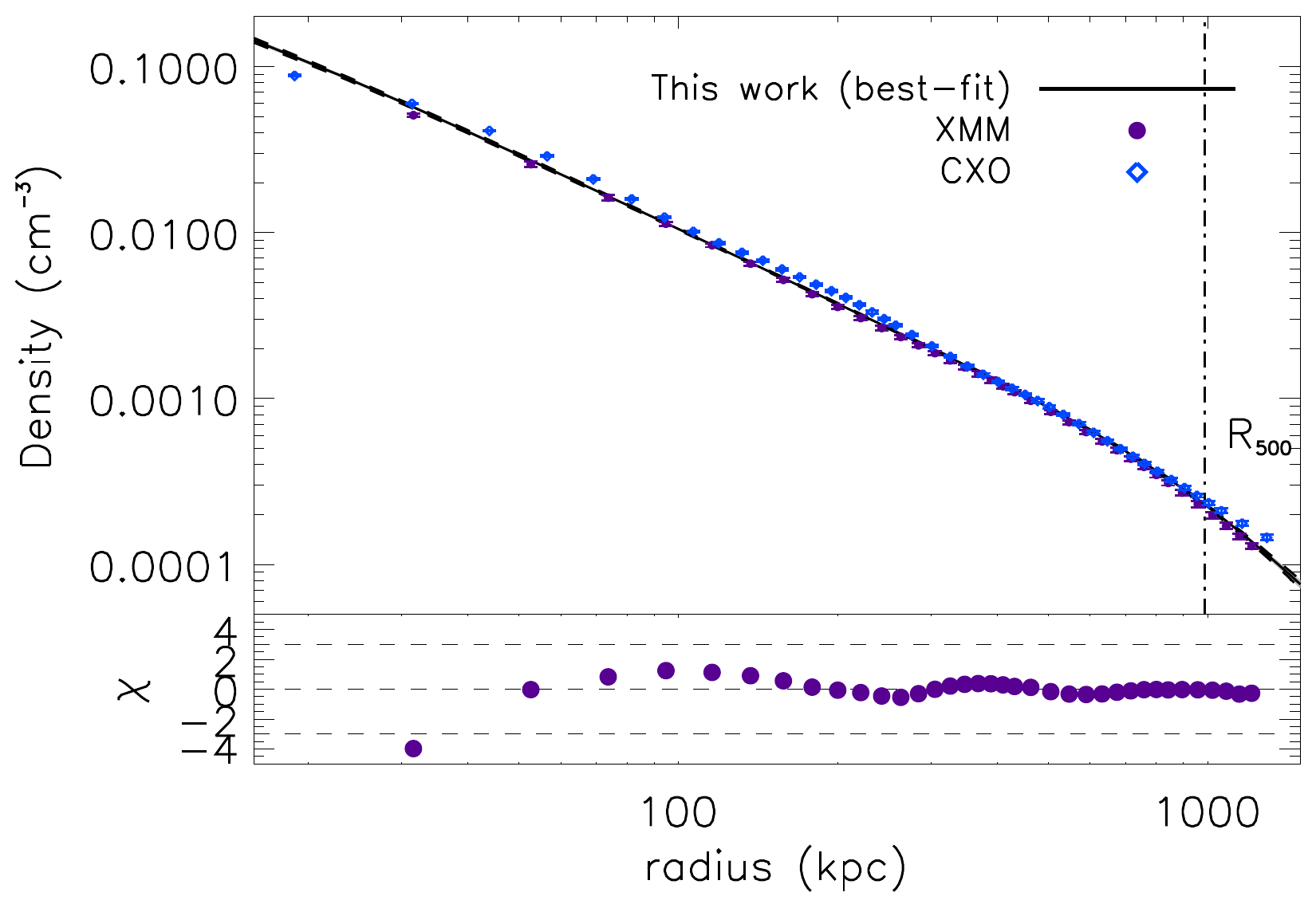}
\caption{\footnotesize MCMC best fit and constraints compared to the NIKA tSZ and XMM-Newton X-ray data used to obtain them. {\bf Left}: Projected tSZ flux density profile from the point source subtracted NIKA data (black dots) and for the best-fit M3 model (red solid line). Uncertainties at 68\% confidence limits are indicated with red dashed lines. In practice,  the tSZ map  is compared to the models and not the profile. {\bf Right}: Deprojected electronic density profile obtained from XMM-Newton (purple dots) and Chandra (CXO for Chandra X-ray Observatory, blue diamonds). The MCMC best-fit model is indicated with a black line, and its uncertainties at 68\% confidence limits are shown with  black dashed lines. The characteristic radius measured from XMM-Newton data, $R_{500} = 986 \pm 10$ kpc, is represented as a vertical dashed line. The residual between XMM-Newton data and the best-fit model normalized by XMM-Newton error bars is also represented on the bottom. The data are well described by the SVM model, except at radii smaller than 30 kpc. See text for details.}
\label{fig:MACSJ1424_MCMC_vs_data}
\end{figure*}

Fig.~\ref{fig:MACSJ1424_MCMC_vs_data} shows a comparison of the MCMC constraints and the data used to obtain them. The left panel presents the best fit and the uncertainties of model M3 compared to the projected flux density profile from the point source-subtracted NIKA data. The data points are correlated across the profile and we accounted for this in the MCMC analysis. The model uncertainties increase significantly in the cluster core because the flux estimate (and associated uncertainty) of the central point source is included in the fit. The model is greater than zero at large scales because the zero level of the map (one of the nuisance parameters included in the fit) is slightly positive. The right panel shows the best fit to the XMM-Newton electronic density and its 68\% confidence limit. We also show the Chandra data for comparison. The X-ray data are well described by the SVM model, except at radii smaller than 30 kpc, where we observe a flattening of the data with respect to the model. However, this region is not probed directly by NIKA and this deviation is not significant in the context of the joint tSZ/X-ray analysis.

\begin{figure*}[h]
\centering
\includegraphics[height=6.3cm]{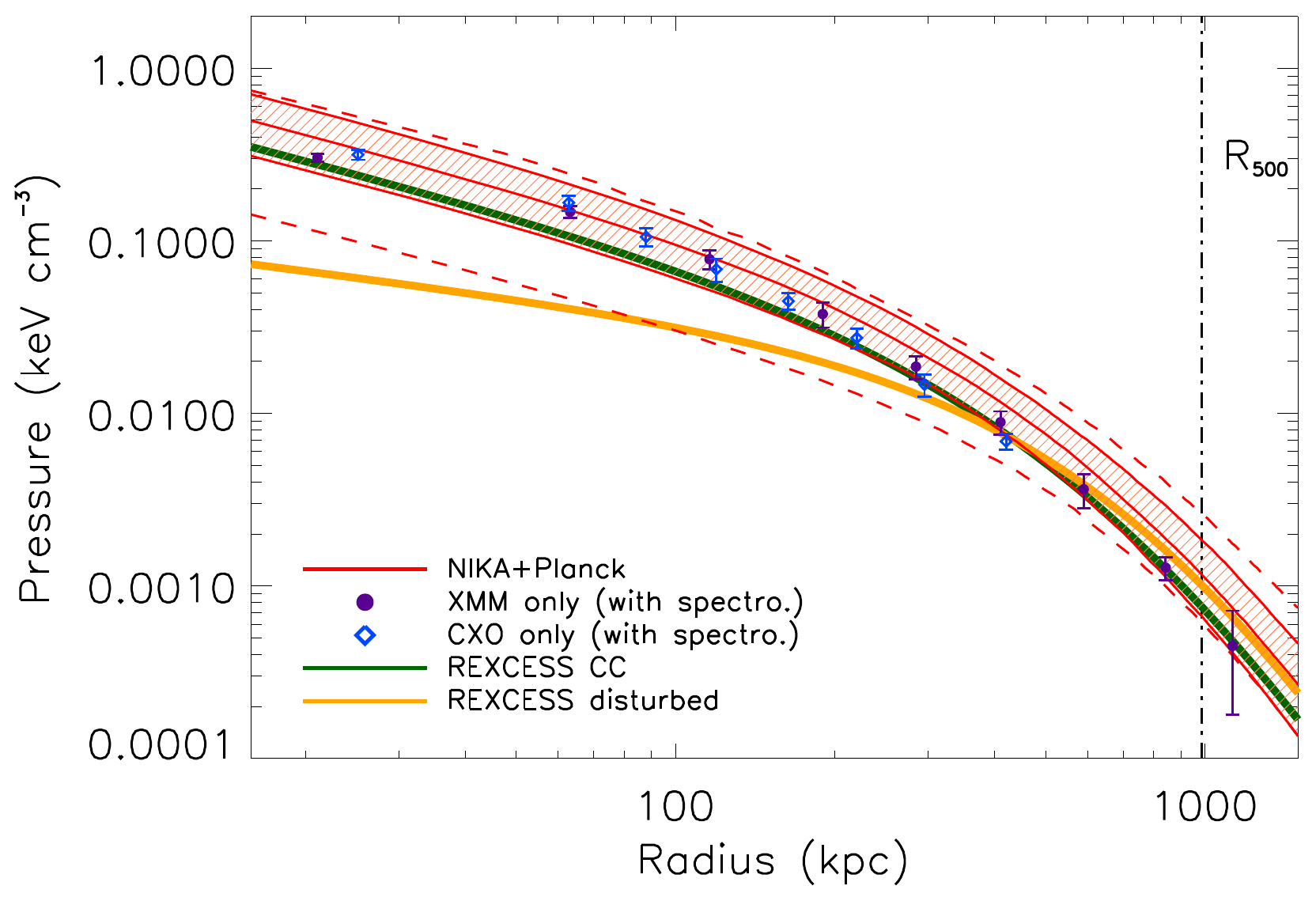}
\includegraphics[height=6.3cm]{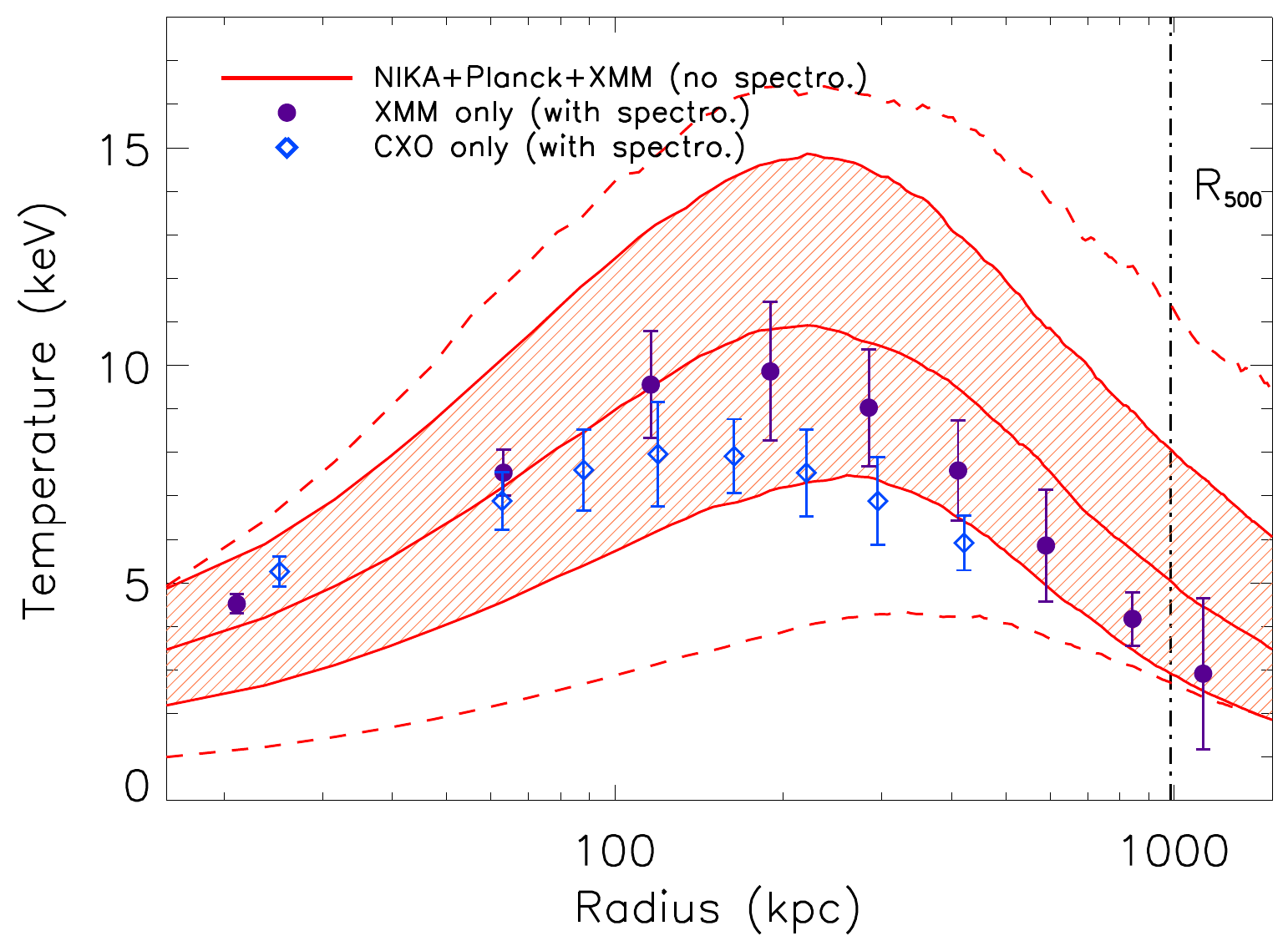}
\includegraphics[height=6.3cm]{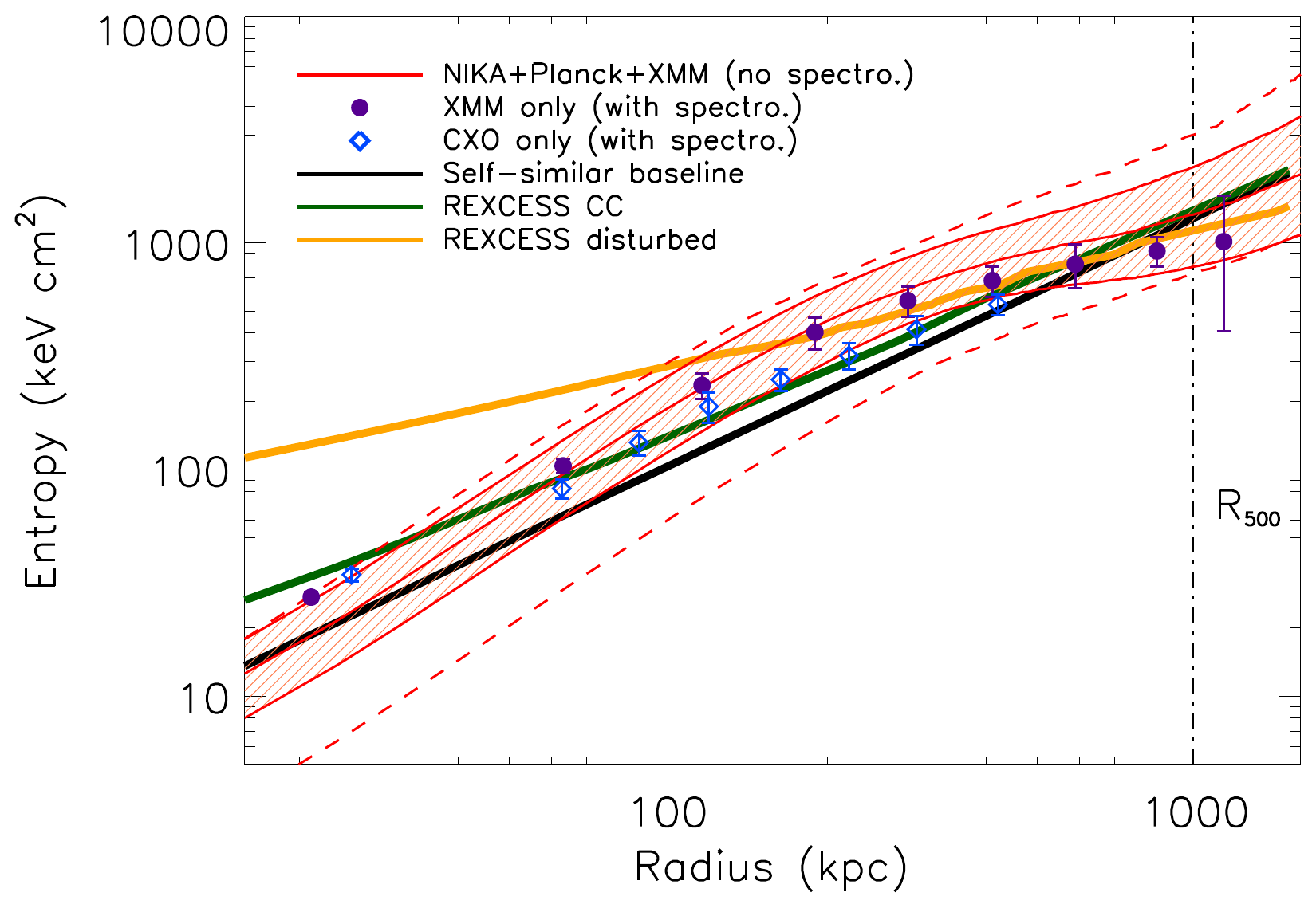}
\includegraphics[height=6.3cm]{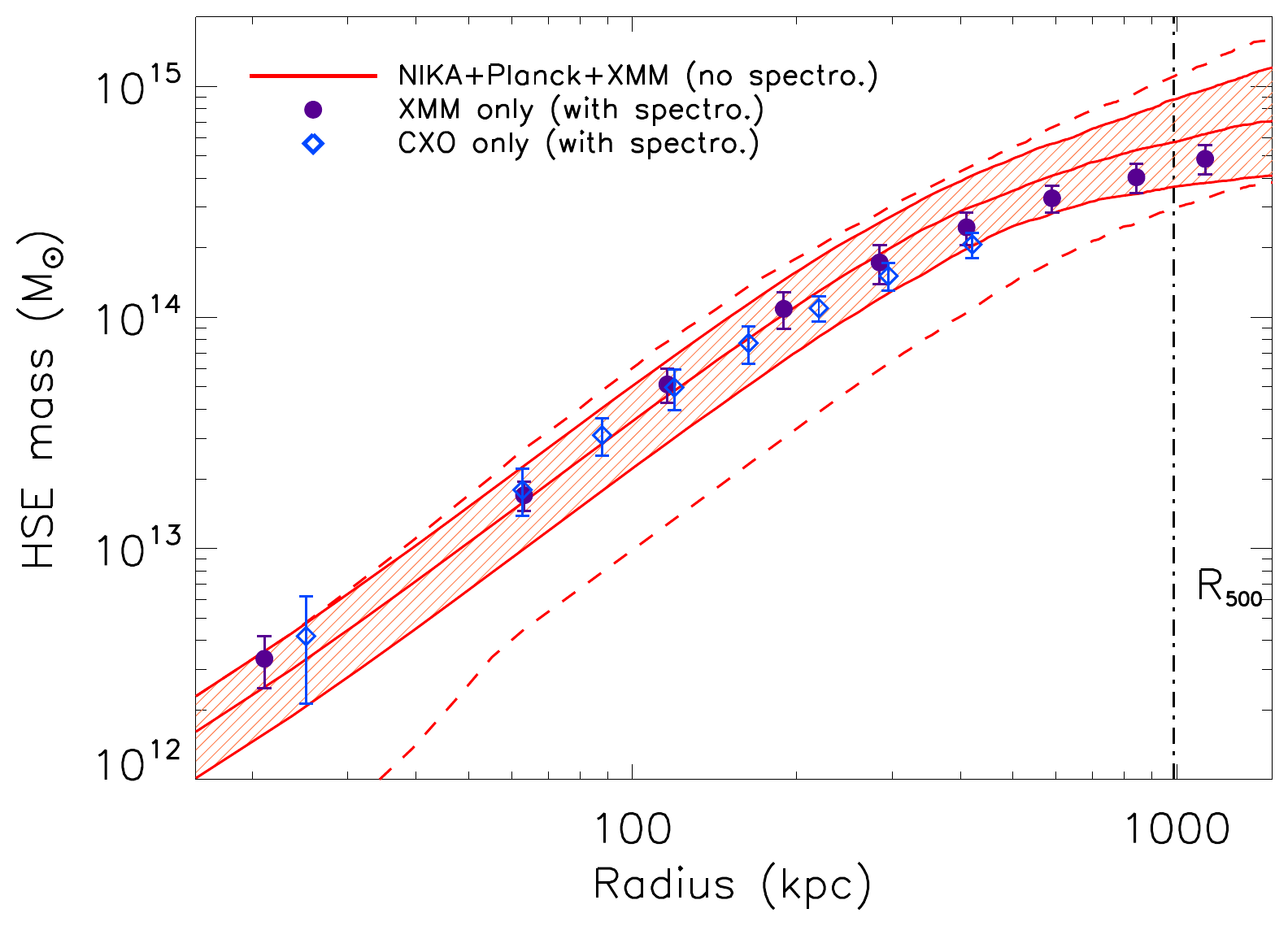}
\caption{\footnotesize MCMC constraints on the deprojected radial profiles of the pressure (top left),  temperature (top right), entropy (bottom left), and  hydrostatic mass (bottom right). The Chandra only and XMM-Newton only measurements are shown with blue diamonds and purple dots, respectively. The red shaded area shows the MCMC 68\% confidence limit from model M2, and the central solid line corresponds to the best-fit model. The two red dashed lines give the upper and lower limits allowed considering all three models: M1, M2, and M3 (see Fig.~\ref{fig:MACSJ1424_pressure_point_source}). In the case of the pressure profile, the constraint is driven by the NIKA and Planck tSZ data. Nevertheless, it also depends slightly on the X-ray derived electronic density since we correct for the relativistic corrections, using the temperature computed from the pressure and density, when comparing a given model to the data. We also show the pressure \citep{arnaud2010} and entropy \citep{pratt2010} profiles of both cool core (green solid line) and morphologically disturbed (orange solid line) clusters based on \rexcess, a representative sample of nearby X-ray clusters. For the entropy profile, the self-similar expectation from nonradiative simulations from \cite{voit2005b} is also represented as a black solid line.}
\label{fig:MACSJ1424_MCMC_tk_profile}
\end{figure*}

The top left panel of Fig.~\ref{fig:MACSJ1424_MCMC_tk_profile} shows the deprojected pressure profile of \mbox{MACS~J1423.8+2404,} as derived from our MCMC analysis of NIKA data and from X-ray measurements only. While deep X-ray observations can only infer the pressure (using the product of the deprojected density and temperature profiles), the tSZ effect directly measures it. The comparison of the two, which are completely independent, is therefore an important cross-check for both constraints. The data points represent measurements obtained from XMM-Newton and Chandra, and the shaded area is the 68\% confidence limit given by model M2. The upper and lower dashed lines give  the upper and lower limits,respectively, from all the tSZ based models M1, M2 and M3, the lower limit being set by M1 and the upper one by M3. The Chandra constraint is slightly more peaked than that from XMM-Newton. In turn, the XMM-Newton constraint is slightly more peaked than that from NIKA, but all profiles are compatible within error bars. These results confirm that the point spread function deconvolution of the XMM-Newton data performs well and that the model extrapolation of the tSZ data accurately describes the pressure profile at the cluster center. At large radii, the Chandra measurement is not available owing to a lower photon count rate  than XMM-Newton, making the two observations highly complementary. The tSZ  pressure profile decreases slightly more slowly than that from XMM-Newton, but the effect is not significant with respect to the uncertainties. 

We compared our measurements to the universal pressure profile obtained by \cite{arnaud2010} using  \rexcess\  \citep{bohringer2007}, a representative sample of nearby clusters. The green and orange solid lines show the mean profiles for the cool core and morphologically disturbed subsamples used by \cite{arnaud2010}, respectively. The normalization (eqn.~13 of \cite{arnaud2010}) was accounted for using the XMM-Newton mass measurement of \mbox{MACS~J1423.8+2404}, and neglecting the mass dependance of the shape of the profile. The observed pressure distribution of  \mbox{MACS~J1423.8+2404} compares well to the mean cool core profile both in shape and amplitude. The mean morphologically disturbed profile is consistent with the observations at large scales, but deviates significantly in the core since it is much shallower than the observations because of the redistribution of the thermal energy from merging events in such clusters. The relaxed dynamical state of \mbox{MACS~J1423.8+2404} is therefore confirmed by its pressure profile. We stress that the point source subtraction is crucial to this result (cf. model M1).

Finally,  our measurement is compatible with the tSZ interferometric results obtained by \cite{bonamente2012} on the same cluster. Using an \cite{arnaud2010} model, \cite{bonamente2012} fit the parameters $P_0$ and $r_p$. Since the two are highly degenerated, we only consider their best-fit pressure profile and we find  that it agrees within 1 $\sigma$ with our models M1 and M2 at scales larger than 100 kpc, and agrees within 2 $\sigma$ over the entire radial range.

The constraints on temperature and entropy, directly related to those from the pressure and the density (see Sect.~\ref{sec:modeling}), are also given in Fig.~\ref{fig:MACSJ1424_MCMC_tk_profile}. The red contours use the tSZ derived pressure and the XMM-Newton derived density, while the data points are X-ray only  measurements based on spectroscopic data. Since both the temperature and entropy are proportional to the pressure, this figure reproduces the comparison of the top left panel of Fig.~\ref{fig:MACSJ1424_MCMC_tk_profile}, but focuses on different thermodynamic quantities. The temperature profile is typical of a cool core cluster with a core temperature of about 4 keV, reaching about 10 keV at a radial distance of 200 kpc. The deprojected XMM-Newton and Chandra temperature profiles are in good agreement over the full radial range. The shapes of the two profiles are also consistent, showing a peak of the temperature at the same radius, even though the Chandra profile is somewhat lower $\sim 1$ keV, contrary to what is generally observed (e.g., \citealt{mahdavi2013} or \citealt{martino2014}).

The spatial distribution of entropy is an important tool for investigating cluster formation as it is related to the structure of the ICM and records the thermodynamical history of the gas \citep[see][for a review]{voit2005}. Numerical simulations only including gravitational processes were used by \cite{voit2005b} to obtain a baseline entropy profile, which is well described by a simple power law of the form $K(r) = 1.32 K_{200} \left(r/R_{200}\right)^{1.1}$. \cite{pratt2010}  used the \rexcess\ sample to investigate the properties of entropy in a representative sample of nearby clusters, observing a central excess in dynamically disturbed systems and a power-law profile in cool core systems (see also results obtained by \cite{cavagnolo2009} using the ACCEPT data). The entropy can be described by a power-law plus constant, $K(r) = K_0+K_{100} \left(r/100 \ h_{70}^{-1} \ {\rm kpc}\right)^{\alpha_{K}}$. Disturbed systems present a higher plateau (up to almost two orders of magnitude) and a shallower slope than cool cores. We used the best-fit (power law plus constant) profiles of \rexcess\ clusters \citep[see Tab.~3 of][]{pratt2010} to derive the median entropy profile of cool core and morphologically disturbed clusters, which we compare to our data in the bottom left panel of Fig.~\ref{fig:MACSJ1424_MCMC_tk_profile}. The baseline profile of \cite{voit2005b} is also included using the normalization described in \cite{pratt2010}, which is appropriate for the mass of \mbox{MACS~J1423.8+2404}. The observed entropy profile compares very well with the median cool core profile from the \rexcess\ sample; it is almost consistent with the single power-law self-similar expectation from nonradiative simulations \citep{voit2005b}. For comparison, the median morphologically disturbed median profile is about one order of magnitude higher than our data at the cluster core. At large scales, both XMM-Newton and the tSZ derived constraints indicate a flattening of the profile, but it is not significant compared to the error bars. As for the pressure profile, the entropy distribution clearly shows that \mbox{MACS~J1423.8+2404} behaves as a typical cool core, and that it is indeed a relaxed system \citep[as discussed in Sect.~\ref{sec:Introduction}, e.g.,][]{kartaltepe2008,limousin2010}. 

We can derive the mass profile assuming hydrostatic equilibrium, as described in Sect.~\ref{sec:modeling}. The bottom right panel of Fig.~\ref{fig:MACSJ1424_MCMC_tk_profile} compares the constraints obtained when combining the X-ray and tSZ data with those obtained using X-ray data only. The constraints are fully compatible at all scales within the 68\% confidence limit. We obtained $R_{500} = 1123^{+328}_{-187}$ kpc, which gives $M_{500} = 6.9^{+6.5}_{-2.9}  \times 10^{14} {\rm M}_{\odot}$, in the case of model M3. These results are fully consistent with those obtained from XMM-Newton data only: ($R_{500} = 986 \pm 10$ kpc and $M_{500} = 4.9 \pm 0.15 \times 10^{14} {\rm M}_{\odot}$), but with larger uncertainties. The total integrated Compton parameter was constrained to be $Y_{\rm tot} = 0.85^{+0.61}_{-0.30} \times 10^{-3}$ arcmin$^2$, in agreement with, but slightly larger than the Planck value from aperture photometry. We summarize the main recovered properties of \mbox{MACS~J1423.8+2404} in Table~\ref{tab:summary}.
\begin{table}[h]
\caption{Summary of the recovered properties of \mbox{MACS~J1423.8+2404}.}
\begin{center}
\begin{tabular}{cc}
\hline
\hline
$M_{500, {\rm X \ only}}$ & $4.9 \pm 0.15 \times 10^{14} {\rm M}_{\odot}$ \\
$M_{500, {\rm X/tSZ}}$ & $6.9^{+6.5}_{-2.9}  \times 10^{14} {\rm M}_{\odot}$\\
$R_{500, {\rm X \ only}}$ & $986 \pm 10$ kpc\\
$R_{500, {\rm X/tSZ}}$ & $1123^{+328}_{-187}$ kpc\\
$\theta_{500, {\rm X \ only}}$ & $2.50 \pm 0.03$ arcmin \\
$\theta_{500, {\rm X/tSZ}}$ & $2.85^{+0.83}_{-0.47}$ arcmin\\
$Y_{\rm tot}$ & $0.85^{+0.61}_{-0.30} \times 10^{-3}$ arcmin$^2$ \\
\hline
\end{tabular}
\end{center}
\label{tab:summary}
\end{table}

The X-ray results present error bars smaller than those derived from NIKA+Planck tSZ data. This is to be expected because both X-ray datasets are particularly exceptional for this cluster (about 30 hours of exposure time for both instruments), while the tSZ datasets have been obtained with 1.47 hours on target. Nevertheless, NIKA provides information that allows the reconstruction of the thermodynamical structure of the cluster to reasonable accuracy, especially when combined with X-ray electronic density.

\section{Conclusions and perspectives}\label{sec:conclusions}
\subsection{Conclusions}
The cluster of galaxies \mbox{MACS~J1423.8+2404} was observed at 150 and 260~GHz using the NIKA camera at the IRAM 30-meter telescope. The target was detected at 150~GHz in only 1.47 hours and the data show evidence for the presence of a contaminating central point source, as expected from previous radio observations. The 260~GHz NIKA band allows the further identification of two sub-millimeter contaminant galaxies.

The NIKA observations were combined with external datasets taken in multiple wavelengths to investigate the cluster morphology and  dynamical state of the gas. The object \mbox{MACS~J1423.8+2404} is a typical cool core, which appears to be relaxed and elliptical. A total of 17 sub-millimeter sources were identified in the NIKA field using Herschel data, and two radio sources were identified from external radio observations.

The fluxes of the two radio sources were extrapolated to the NIKA bands by fitting a power-law SED to external photometric measurements. The sub-millimeter sources were extrapolated to the NIKA bands by modeling their SED with a gray-body spectrum and by jointly fitting the Hershel and NIKA photometric data. This work shows the excellent complementarity between the two instruments in terms of frequency coverage and angular resolution. A comparison of the flux density derived directly from NIKA data  shows  good agreement with the expectation from extrapolated point source fluxes at the same frequencies. The contamination from point sources can, therefore, be estimated and subtracted from the tSZ map if external data are available at complementary wavelengths.

The pressure profile of \mbox{MACS~J1423.8+2404} was then measured by modeling and jointly fitting the NIKA and Planck (i.e., $Y_{\rm tot}$) data. We have evaluated the impact of the point source contamination on its reconstruction. We find that neglecting the sources leads to an overall underestimate of the profile, which is small compared to the error bars in the case of the available data, but  could become significant when deeper tSZ observations are available. Moreover, in the case of the presence of a point source at the cluster center with no strong priors on its flux, the inner pressure profile remains poorly constrained by the data. Consequently, the morphological characteristics of the cluster could be misunderstood in such cases. We used XMM-Newton and Chandra X-ray observations to derive the cluster thermodynamical radial distributions. The deprojected electronic density has been combined with the tSZ data, providing a measurement of the pressure, and used to infer the temperature and the entropy profiles. The X-ray only (including spectroscopy) and the tSZ+X-ray (without spectroscopy) constraints are consistent within their uncertainties, and confirm that \mbox{MACS~J1423.8+2404} is a relaxed cool core cluster.

\subsection{Prospect for NIKA2 observations}
In the coming years, the New IRAM Kids Array 2 \citep[NIKA2][]{monfardini2014} will be used to map the tSZ signal from clusters of galaxies with an 18 arcsec angular resolution at 150~GHz. The NIKA2 camera\footnote{\url{http://ipag.osug.fr/nika2/Welcome.html}} is the next generation continuum instrument for the IRAM (Institut de Radio Astronomie Millim\'etrique) 30-meter telescope near Granada, Spain. It consists of a dual-band camera made of about 5000 kinetic inductance detectors sampling a 6.5 arcmin field of view, observing the sky at 150 and 260~GHz. The NIKA2 camera was installed during the autumn 2015 and will be commissioned  the following winter. The observations presented in this paper and the present analysis are part of the NIKA tSZ pilot study aimed at characterizing the potential outcomes of future NIKA2 tSZ observations.

Clusters are crowded environments, therefore, in high angular resolution tSZ observations like those of NIKA2 we expect the detection of a large number of point sources, as demonstrated in \cite{adam2013}, \cite{adam2014} and in the present work. These sources are a contaminant in the context of tSZ studies. We have shown here that their contribution can be estimated and accounted for if external data are available. Bright sub-millimeter point sources should not be a problem by themselves because they can be identified at 260~GHz and marginalized over at 150~GHz. They can even be masked without removing the information on the tSZ profile as long as they do not coincide with the cluster center. Fainter, but more numerous, sub-millimeter objects will require higher frequency data to be removed, as carried out in this paper. As we expect a large number of sources below the noise, but contributing to the overall signal, the sub-millimeter galaxy population could be one of the limiting factors of future deeper tSZ observations. Radio point sources can also be problematic. Unless they are as strong as the tSZ signal, they can only be identified using external data. They are generally correlated with the central BCG and will prevent NIKA2 from putting any constraints on the core pressure distribution unless their SED is well measured at lower frequencies. If no external data are available, the removal of the contaminants will require the use of extra high angular resolution observations obtained with interferometers at similar frequencies, such as the IRAM NOrthern Extended Millimeter Array (NOEMA)\footnote{\url{http://iram-institute.org/EN/noema-project.php}}. 

Point sources around clusters are also interesting in themselves and the high complementarity between Herschel and NIKA could be used, for example, to search for distant lensed galaxies \citep[see, for example,][]{egami2010} or to study star formation at high redshift. This should be an extra outcome of the NIKA2 tSZ observations.

The present work has also shown how X-ray and tSZ data can be used to constrain the temperature and entropy profiles of galaxy clusters independently from X-ray spectroscopy using  density and pressure profiles. Moving to high redshift, X-ray observations become time expensive and high-quality X-ray mapping becomes challenging because of  redshift dimming. In this context, the results we obtained show that we are able to constrain the ICM thermodynamics with a good accuracy by combining resolved NIKA tSZ observations and X-ray mapping. This will be particularly important for distant clusters observed with NIKA2, for which we will not benefit from deep X-ray observations tuned for spatially resolved X-ray spectroscopy, and where only a mean temperature can be obtained from X-ray data. In addition, X-ray constraints on the cluster thermodynamics directly depend on the energy calibration of the instrument, which can be source of discrepancy. NIKA tSZ data offer a third independent measurement, independent from X-ray spectroscopy, providing a cross-check of the temperature measurement of clusters. This can help reducing the systematic effects affecting all these observations, which are essential for mass estimates when using clusters as a cosmological probe.

\begin{acknowledgements}
We are thankful to the anonymous referee for useful comments that helped improve the quality of the paper.
We thank Marco De Petris for useful comments.
We would like to thank the IRAM staff for their support during the NIKA campaign.
We thank Karin Dassas for helping us obtain the Herschel data.
This work has been partially funded by the Foundation Nanoscience Grenoble, the ANR under the contracts "MKIDS" and "NIKA". 
This work has been partially supported by the LabEx FOCUS ANR-11-LABX-0013. 
This work has benefited from the support of the European Research Council Advanced Grant ORISTARS under the European Union's Seventh Framework Program (Grant Agreement no. 291294).
This work has benefited from the support of the European Research Council Advanced Grant M2C under the European Union‚Äôs Seventh Framework Programme (Grant Agreement no. 340519).
The NIKA dilution cryostat was designed and built at the Institut N\'eel. In particular, we acknowledge the crucial contribution of the Cryogenics Group and, in particular Gregory Garde, Henri Rodenas, Jean Paul Leggeri, and Philippe Camus. 
We acknowledge the use of the IGLO-HESIOD service from the Integrated Data \& Operation Center (IDOC). Support for IDOC is provided by CNRS \& CNES. 
R. A. would like to thank the ENIGMASS French LabEx for funding this work. 
B. C. acknowledges support from the CNES postdoctoral fellowship program. 
A. R. acknowledges support from the CNES doctoral fellowship program. 
E. P. acknowledges the support of the French Agence Nationale de la Recherche under grant ANR-11-BD56-015.
\end{acknowledgements}

\bibliography{biblio_macsj1424}
\end{document}

%% file: listeauthors.tex
\author{R.~Adam\inst{\ref{inst1}, \ref{inst16}}\thanks{Corresponding author: R\'emi Adam, \url{remi.adam@oca.eu}}
\and B.~Comis\inst{\ref{inst1}}
\and I.~Bartalucci\inst{\ref{inst4}}
\and A.~Adane\inst{\ref{inst2}}
\and P.~Ade\inst{\ref{inst3}}
\and P.~Andr\'e\inst{\ref{inst4}}
\and M.~Arnaud\inst{\ref{inst4}}
\and A.~Beelen\inst{\ref{inst5}}
\and B.~Belier\inst{\ref{inst6}}
\and A.~Beno\^it\inst{\ref{inst7}}
\and A.~Bideaud\inst{\ref{inst3}}
\and N.~Billot\inst{\ref{inst8}}
\and O.~Bourrion\inst{\ref{inst1}}
\and M.~Calvo\inst{\ref{inst7}}
\and A.~Catalano\inst{\ref{inst1}}
\and G.~Coiffard\inst{\ref{inst2}}
\and A.~D'Addabbo\inst{\ref{inst7}, \ref{inst14}}
\and F.-X.~D\'esert\inst{\ref{inst9}}
\and S.~Doyle\inst{\ref{inst3}}
\and J.~Goupy\inst{\ref{inst7}}
\and B.~Hasnoun\inst{\ref{inst5}}
\and I.~Hermelo\inst{\ref{inst8}}
\and C.~Kramer\inst{\ref{inst8}}
\and G.~Lagache\inst{\ref{inst15}}
\and S.~Leclercq\inst{\ref{inst2}}
\and J.-F.~Mac\'ias-P\'erez\inst{\ref{inst1}}
\and J.~Martino\inst{\ref{inst5}}
\and P.~Mauskopf\inst{\ref{inst3}, \ref{inst13}}
\and F.~Mayet\inst{\ref{inst1}}
\and A.~Monfardini\inst{\ref{inst7}}
\and F.~Pajot\inst{\ref{inst5}}
\and E.~Pascale\inst{\ref{inst3}}
\and L.~Perotto\inst{\ref{inst1}}
\and E.~Pointecouteau\inst{\ref{inst10}, \ref{inst11}}
\and N.~Ponthieu\inst{\ref{inst9}}
\and G.W.~Pratt\inst{\ref{inst4}}
\and V.~Rev\'eret\inst{\ref{inst4}}
\and A.~Ritacco\inst{\ref{inst1}}
\and L.~Rodriguez\inst{\ref{inst4}}
\and G.~Savini\inst{\ref{inst12}}
\and K.~Schuster\inst{\ref{inst2}}
\and A.~Sievers\inst{\ref{inst8}}
\and S.~Triqueneaux\inst{\ref{inst7}}
\and C.~Tucker\inst{\ref{inst3}}
\and R.~Zylka\inst{\ref{inst2}}}

\institute{
Laboratoire de Physique Subatomique et de Cosmologie, Universit\'e Grenoble-Alpes, CNRS/IN2P3, 53, rue des Martyrs, Grenoble, France
  \label{inst1}
  \and
  Laboratoire Lagrange, Universit\'e C\^ote d'Azur, Observatoire de la C\^ote d'Azur, CNRS, Blvd de l'Observatoire, CS 34229, 06304 Nice cedex 4, France
  \label{inst16}
  \and
Laboratoire AIM, CEA/IRFU, CNRS/INSU, Universit\'e Paris Diderot, CEA-Saclay, 91191 Gif-Sur-Yvette, France 
  \label{inst4}
\and
Institut de RadioAstronomie Millim\'etrique (IRAM), Grenoble, France
  \label{inst2}
\and
Astronomy Instrumentation Group, University of Cardiff, UK
  \label{inst3}
\and
Institut d'Astrophysique Spatiale (IAS), CNRS and Universit\'e Paris Sud, Orsay, France
  \label{inst5}
\and
Institut d'Electronique Fondamentale (IEF), Universit\'e Paris Sud, Orsay, France
  \label{inst6}
\and
Institut N\'eel, CNRS and Universit\'e de Grenoble, France
  \label{inst7}
\and
Institut de RadioAstronomie Millim\'etrique (IRAM), Granada, Spain
  \label{inst8}
\and
Dipartimento di Fisica, Sapienza Universit\`a di Roma, Piazzale Aldo Moro 5, I-00185 Roma, Italy
  \label{inst14}
\and
Institut de Plan\'etologie et d'Astrophysique de Grenoble (IPAG), CNRS and Universit\'e de Grenoble, France
  \label{inst9}
    \and
Aix Marseille Universit\'e, CNRS, LAM (Laboratoire d'Astrophysique de Marseille) UMR 7326, 13388, Marseille, France
  \label{inst15}
\and
School of Earth and Space Exploration and Department of Physics, Arizona State University, Tempe, AZ 85287
  \label{inst13}
\and
Universit\'e de Toulouse, UPS-OMP, Institut de Recherche en Astrophysique et Plan\'etologie (IRAP), Toulouse, France
  \label{inst10}
\and
CNRS, IRAP, 9 Av. colonel Roche, BP 44346, F-31028 Toulouse cedex 4, France 
  \label{inst11}
\and
University College London, Department of Physics and Astronomy, Gower Street, London WC1E 6BT, UK
  \label{inst12}
}